\documentclass[aps,pra,twocolumn,showpacs,superscriptaddress,showkeys,nofootinbib,longbibliography]{revtex4-1}

\usepackage[utf8]{inputenc}
\usepackage[T1]{fontenc}

\usepackage{multirow}	
\usepackage{enumerate}	
\usepackage{Tabbing}	
\usepackage{relsize}	
\usepackage{textcomp}	
\usepackage{soul}	

\usepackage{amsmath}
\usepackage{amssymb}
\usepackage{bm}		

\usepackage{hyperref}
\usepackage{url}
\usepackage[capitalise]{cleveref}	

\usepackage{dcolumn}	
\usepackage{longtable}	
\usepackage{booktabs}	

\usepackage[pdftex]{graphicx}
\usepackage{epstopdf}	
\usepackage{rotating}	
\usepackage{float}	

\usepackage{listings}
\usepackage{color}
\usepackage{textcomp}
\usepackage[percent]{overpic}

\definecolor{listinggray}{gray}{0.9}
\definecolor{lbcolor}{rgb}{0.9,0.9,0.9}

\usepackage[dvipsnames]{xcolor}

\newcommand{\ket}[1]{\ensuremath{\left|#1\right\rangle}}
\newcommand{\average}[1]{\ensuremath{\left\langle#1\right\rangle}}
\newcommand{\textav}[1]{\ensuremath{\langle#1\rangle}}
\newcommand{\bracket}[2]{\ensuremath{\left\langle#1 \vphantom{#2}\right| \left. #2 \vphantom{#1}\right\rangle}}
\newcommand{\matrixel}[3]{\ensuremath{\left\langle #1 \vphantom{#2#3} \right| #2 \left| #3
\newcommand{\angstrom}{\text{\normalfont\AA}}
\vphantom{#1#2} \right\rangle}}

\newcommand{\AN}[2]{
\ensuremath{\hat{c}^{} _{{#1}
\ifnum#2=1 \uparrow
\else
\ifnum#2=-1 \downarrow
\else
\ifnum#2=2 \sigma
\else
\ifnum#2=-2 \bar{\sigma}
\else
\ifnum#2=3 \sigma '
\fi
\fi
\fi
\fi
\fi
}}
}
\newcommand{\CR}[2]{
\ensuremath{\hat{c}^\dagger _{{#1}
\ifnum#2=1 \uparrow
\else
\ifnum#2=-1 \downarrow
\else
\ifnum#2=2 \sigma
\else
\ifnum#2=-2 \bar{\sigma}
\else
\ifnum#2=3 \sigma '
\fi
\fi
\fi
\fi
\fi
}}
}
\newcommand{\NUM}[2]{
\ensuremath{\hat{n}^{} _{{#1}
\ifnum#2=1 \uparrow
\else
\ifnum#2=-1 \downarrow
\else
\ifnum#2=2 \sigma
\else
\ifnum#2=-2 \bar{\sigma}
\else
\ifnum#2=3 \sigma '
\fi
\fi
\fi
\fi
\fi
}}
}

\newcommand{\mc}[3]{\multicolumn{#1}{#2}{#3}}

\renewcommand{\vec}[1]{\ensuremath{\mathbf{#1}}}

\definecolor{newRed}{RGB}{200,0,0}

\definecolor{newGreen}{RGB}{0,100,0}

\begin{document}


\title {Metallization of solid molecular hydrogen in two dimensions:\texorpdfstring{\\}{} Mott-Hubbard-type transition}

\author{Andrzej Biborski}
\email {andrzej.biborski@agh.edu.pl}
\affiliation {Academic Centre for Materials and Nanotechnology,
AGH University of Science and Technology,
al. Mickiewicza 30, PL-30-059 Krak\'ow, Poland}

\author{Andrzej P. K\k{a}dzielawa}
\email{kadzielawa@th.if.uj.edu.pl}
\affiliation{Marian Smoluchowski Institute of Physics, Jagiellonian University, 
ulica \L{}ojasiewicza 11, PL-30-348 Krak\'ow, Poland}

\author{J\'{o}zef Spa\l{}ek}
\email{jozef.spalek@uj.edu.pl}
\affiliation{Marian Smoluchowski Institute of Physics, Jagiellonian University, 
ulica \L{}ojasiewicza 11, PL-30-348 Krak\'ow, Poland}

\date{\today}

\begin{abstract}
 We analyze the pressure-induced metal-insulator transition in a two-dimensional vertical stack of $H_2$ molecules in (\emph{x-y}) plane, and show that it represents a striking example of the Mott-Hubbard-type transition. Our combined exact diagonalization approach, formulated and solved in the second quantization formalism, includes also  simultaneous \emph{ab initio} readjustment of the single-particle wave functions, contained in the model microscopic parameters. The system is studied as a function of applied side force (generalized pressure), both in the $H_2$-molecular and $H$-quasiatomic states. Extended Hubbard model is taken at the start, together with longer-range electron-electron interactions incorporated into the scheme. The stacked molecular plane transforms discontinuously into a (quasi)atomic state under the applied force via a two-step transition: the first between molecular insulating phases and the second from the molecular to the quasiatomic metallic phase. No quasiatomic insulating phase occurs. All the transitions are accompanied by an abrupt changes of the bond length and the intermolecular distance (lattice parameter), as well as by discontinuous changes of the principal electronic properties, which are characteristic of the Mott--Hubbard transition here associated with the jumps of the predetermined equilibrium lattice parameter and the effective bond length. The phase transition can be interpreted in terms of the solid hydrogen metallization under pressure exerted by e.g., the substrate covered with a monomolecular $H_2$ film of the vertically stacked molecules. Both the Mott and Hubbard criteria at the insulator to metal transition are discussed.
 
\end{abstract}


\maketitle

\section{Motivation}
\label{sec:motivation}
Hydrogen is the first and the simplest of elements in the Periodic Table, with an elementary structure of the energy levels. Also, the $H_2$ molecule represents the testing ground of quantum-mechanical methods \cite{Kolos,Pachucki}. This elementary nature of the atomic or molecular energy levels transforms into an involved manifold of states and available energies as exemplified by the abundance of their condensed liquid and solid phases \cite{Dalladay-Simpson,Dzyabura}.
The resultant phase diagram is complex, and the catalogue of observed phases - especially of the solid ones - steadily increases \cite{Dalladay-Simpson,Howie}. The lack of clarity concerning their crystal structure in many cases is intimately connected with an incomplete insight into their electronic properties. 
However, it has been unclear until very recently \cite{Dias} whether the solid-hydrogen atomic and metallic phase may indeed exist. Nonetheless, the detailed nature of this transition from an insulating molecular phase to the (quasi-)metallic atomic state, is still under debate \cite{McMinis,Drummond}, starting from the historic paper by Wigner and Huntington \cite{Wigner}. Once confirmed \cite{EremetsDrozdov}, the recent work \cite{Dias} would represent a decisive step in achieving our understanding of the metallization of molecular hydrogen both experimentally and theoretically. The fundamental question is whether this transition is of the \textbf{Mott-Hubbard type}, i.e., driven by the interelectronic correlations \cite{Mott,Gebhard} or is it in class of general \textbf{dielectric--metal transition} driven simply by the formation of overlapping bands under strong pressure \cite{Landau,Landau2}. The principal purpose of the present paper is to provide an affirmative answer to the former possibility, albeit limited to a two-dimensional situation.

Our discussion of the problem is based on an original method of approach, so it is proper to sketch first the context of the current theoretical methods applied. Many, if not most of the attempts performed up to now are based on the\emph{ Density Functional Theory} (DFT) approach. However, as it was reported by Azadi et al.\cite{Azadi}, the results coming from the DFT are often ambiguous and depend strongly on a selection of the form of the \emph{correlation-exchange } potential. Furthermore, obtaining a proper asymptotic behavior (i.e., the value of the dissociation energy) for the $H_2$ molecule in the large-intermolecular-separation limit is also questionable, or at least not straightforward within that approach. Whereas a proper description of the dissociation is crucial for the proper description of the metallization, as well as for the molecular crystal stabilization by taking into account the long-range London dispersion forces, a proper account of the \emph{electron-electron} correlations is regarded by us as equally important. Also, the DFT-based methods such as LDA+U, LDA+DMFT suffer from the so-called \emph{double-counting} problem, making their usability questionable for these systems, where the interelectronic correlations play the crucial role, particularly for low-dimensional systems. In this work we apply a specific, in principle \emph{rigorous method} called the EDABI (\textbf{E}xact \textbf{D}iagonalization \textbf{Ab}-\textbf{I}nitio method) which allows to surpass the last difficulty \cite{Kadzielawa,Biborski,Kadzielawa2,Spalek}. However, the scope of this work is more general. Namely, we treat carefully the interelectronic interactions in the second quantization scheme and concurrently readjust variationally the single-particle wave functions, contained in the microscopic parameters, when constructing the resultant system correlated state. This method of approach thus inverts the order of executing the whole program of determining the electronic properties by diagonalizing the Hamiltonian including interactions in the second-quantization language and determining concomitantly the single-particle wave functions. Also, the present work is an essential extension of our recent communication \cite{Kadzielawa} on quasi-one-dimensional hydrogen ladder to the two-dimensional ($2D$) situation. Namely, we provide details of both the general methodological aspects of our approach and the concrete results for the $2D$ stack of $H_2$ molecules (depicted schematically in Fig.~\ref{fig:plane_model}). We map the whole problem onto the \emph{extended Hubbard} model in which we additionally include the long-range (intermolecular) nature of interaction between electrons. From this point of view, we investigate the physical properties in an \emph{exact manner} within the decomposition of the whole system into periodic units, each containing $4$ molecules. In particular, we focus on the Mott-Hubbard physics of the system by generalizing it to the situation when an insulating and \emph{diamagnetic} molecular $2D$ solid transforms into a paramagnetic atomic and metallic bilayer of $H$ atoms.

The structure of the paper is as follows. In the following Section we provide description of the applied methodology and detail the model. Next, we discuss the phase transition induced by an external side force (effective pressure) and relate it to that of the Mott--Hubbard transition for correlated systems. Finally, we discuss a possible extension of the method to the three -- dimensional ($3D$) systems which represent a final, not yet achieved goal within our method.

\section{Method: Exact Diagonalization - Ab Initio approach (EDABI)}

Our methodology of approach is based on the variational approach which is an extension of the elaborated earlier in our group \textbf{E}xact \textbf{D}iagonalization \textbf{Ab} \textbf{I}nitio (EDABI) scheme in the following manner \cite{Kadzielawa,Biborski,Kadzielawa2,Spalek}. EDABI combines both the first- and the second-quantization schemes. What is fundamentally important, in this work we go both beyond the parametrized-model methodology \cite{Hubbard,Fazekas,Fulde} and put the emphasis first on the interelectronic correlations and simultaneously renormalize the single-particle wavefunctions when constructing the resultant correlated state. To achieve this goal we start with the general electronic Hamiltonian in a second-quantization form representing an interacting system of fermions \cite{Fetter}, i.e.,
\begin{align}
 \label{eq:hamiltonian_general}
 \hat{\mathcal{H}} =& \sum_\sigma \int d^3r \hat{\Psi}^{{\dagger}}_\sigma ( \vec{r} ) \hat{\mathcal{H}}_1 ( \vec{r} ) \hat{\Psi}^{\phantom{\dagger}}_\sigma ( \vec{r} ) \\\notag
		  &+ \frac{1}{2} \sum_{\sigma \sigma'} \iint d^3r d^3r' \hat{\Psi}^{{\dagger}}_\sigma ( \vec{r} ) \hat{\Psi}^{{\dagger}}_{\sigma'} ( \vec{r}' ) \hat{V} ( \vec{r} - \vec{r}' ) \hat{\Psi}^{\phantom{\dagger}}_{\sigma'} ( \vec{r}' ) \hat{\Psi}^{\phantom{\dagger}}_\sigma ( \vec{r} ).
\end{align}
Hamiltonians in the first (canonical) quantization are for single ($\hat{\mathcal{H}}_1$) and pair of particles ($\hat{V} ( \vec{r} - \vec{r}' )$) respectively. $ \hat{\Psi}^{\phantom{\dagger}}_\sigma ( \vec{r} )$ and $\hat{\Psi}^{{\dagger}}_\sigma ( \vec{r} )$ are the field operator and its adjoint, respectively. By introducing fermionic creation and anihilation operators ($\CR{i}{2}$ and $\AN{i}{2}$), conforming the usual anticommutation relations
\begin{align}
 \{ \CR{i}{2}, \CR{j}{3} \} \equiv \{ \AN{i}{2}, \AN{j}{3} \} \equiv 0 \ \ \ \text{and} \ \ \{ \CR{i}{2}, \AN{j}{3} \} \equiv \delta_{ij}\delta_{\sigma \sigma'},
\end{align}
where $\sigma$ denotes spin variable, the field operators can be represented by an expansion in the creation(anihilation) operators, weighted with the amplitudes which represent single-particle wave functions $\{w_i ( \vec{r} )\}$ forming a complete and orthogonal basis in the Hilbert space, i.e.,
\begin{align}
 \label{eq:field_opers}
 \hat{\Psi}^{\phantom{\dagger}}_\sigma ( \vec{r} ) = \sum_{i} w_i ( \vec{r} ) \AN{i}{2},\quad  \hat{\Psi}^{\dagger}_\sigma ( \vec{r} ) = \sum_{i} w_i ( \vec{r} ) \CR{i}{2}.
\end{align}

Hamiltonian \eqref{eq:hamiltonian_general} consists of one--electron part associated with the Hamiltonian for a single particle

\begin{align}
 \label{eq:one-body-ham}
 \hat{\mathcal{H}}_1 ( \vec{r} ) &\overset{a.u.}{=} -\nabla^{2} - \sum_{i=1}^{N_S}\frac{2}{|\vec{R_i} - \vec{r}|},
\end{align}
where $\vec{R_i}$ refers to the coordination of atomic centre and $N_S$ is the number of sites,
and of the \emph{electron-electron} interaction part
\begin{align}
 \label{eq:two-body-ham}
 \hat{V}( \vec{r} - \vec{r}' ) &\overset{a.u.}{=} \frac{2}{|\vec{r} - \vec{r'}|}.
\end{align}
In both equations we used atomic units (a.u).
Combining equations \eqref{eq:hamiltonian_general} and \eqref{eq:field_opers} leads to the Hamiltonian expressed in the language of creation and annihilation operators in the usual form
\begin{align}
 \label{eq:hamiltonian}
 \hat{\mathcal{H}} = \sum\limits_{ij}\sum\limits_{\sigma}t_{ij}\CR{i}{2}\AN{j}{2} +\sum\limits_{ijkl}\sum\limits_{\sigma,\sigma'}V_{ijkl}\CR{i}{2}\CR{j}{3}\AN{l}{3}\AN{k}{2},
\end{align}
where $t_{ij}$ and $V_{ijkl}$ are one- and two-electron interaction parameters defined as
\begin{subequations}
\label{eq:microscopicP}
\begin{align}
 \label{eq:one-body}
 t_{ij} &\equiv \matrixel{w_{i}(\vec{r})}{\hat{\mathcal{H}}_1}{w_j(\vec{r})} \\\notag
     &= \int d^3r \ w_i^*(\vec{r}) \hat{\mathcal{H}}_1 (\vec{r}) w_j(\vec{r}) , \\
 \label{eq:two-body}
 V_{ijkl} &\equiv \matrixel{w_i(\vec{r}) w_j(\vec{r'})}{\hat{V}}{w_k(\vec{r})w_l(\vec{r}')} \\\notag
	  &= \iint d^3r d^3r' \ w_i^*(\vec{r}) w_j^*(\vec{r}') \hat{V} ( \vec{r} - \vec{r}' ) w_k(\vec{r}) w_l(\vec{r}') .
\end{align}
\end{subequations}

In the computationally tractable scheme expansion \eqref{eq:field_opers} is truncated, i.e., the sum in \eqref{eq:field_opers} is assumed as \emph{finite}. Additionally, the functions $\{w_i{(\vec{r})}\}$ in the expansion have their own, or may be supplied with, \emph{internal} parameters $\{\lambda\}$, in addition to the quantum numbers characterized by the set $\{i\}$. These parameters might be used in the variational procedure to optimize the finite-size basis composing an approximate form of $\hat{\Psi}^{\phantom{\dagger}}_\sigma$, in the correlated state, i.e., 

\begin{align}
 \label{eq:field_opers_approx}
 \hat{\Psi}^{\phantom{\dagger}}_\sigma ( \vec{r} ) \approx \sum_{i}^{M} w_i^{(\{\lambda\})} ( \vec{r} ) \AN{i}{2},
\end{align}
where $M$ is a finite number. In that situation,the integrals defined in \eqref{eq:microscopicP} depend also on $\{\lambda\}$ and, in effect, we obtain a \emph{trial Hamiltonian} $\hat{\mathcal{H}}^{(\{\lambda\})}$, for which we solve eigenequation (in our case by means of the Lanczos diagonalization method) for the many-electron problem, i.e.,

\begin{align}
 \label{eq:eigen_problem}
 \hat{\mathcal{H}}^{(\{\lambda\})}\ket{\Psi_{T}^{(\{\lambda\})}} = E_{T}^{(\{\lambda\})}\ket{\Psi_{T}^{(\{\lambda\})}}, 
\end{align}
where $E_T^{(\{\lambda\})}$ is a trial eigenvalue related to the $\ket{\Psi_{T}^{(\{\lambda\})}}$ trial many-body state. The variational procedure relies on finding the minimum of $E_T^{(\{\lambda\})}$ with respect to $\{\lambda\}$. Accordingly, the procedure is limited to relatively small systems, containing typically over a dozen electrons and corresponding to them single--particle states, providing an \emph{exact} solution, at least in principle. As we have shown previously \cite{Biborski}, the calculation of integrals \eqref{eq:microscopicP} can be expensive in terms of the computational time. Below we provide the procedure of evaluating them.
Note that the diagonal hopping element $t_{ii}$, i.e., the single-particle atomic energy is here also important as we discuss the system evolution with pressure which alters also the atomic energy.

\section{Starting system: Two-dimensional stack of \texorpdfstring{$H_2$}{H2} molecules }
\label{sec:system}
\begin{figure}
 \includegraphics[width=\linewidth]{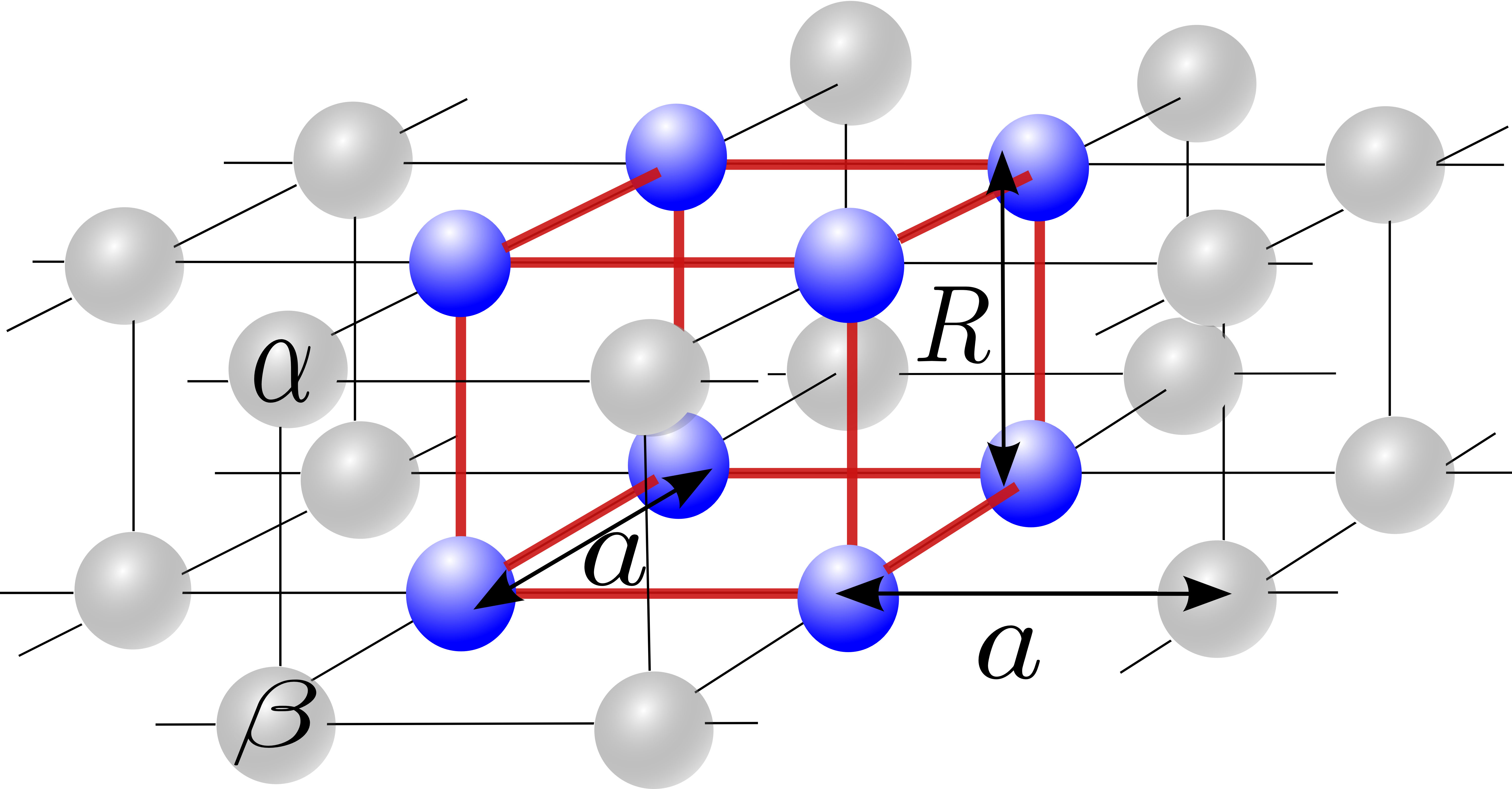}
 \caption{Schematic representation of stacked vertically $H_2$ molecular $2D$ layer forming square lattice. The bond length and the intermolecular distance are marked by $R$ and $a$, respectively. There are eight
 atoms in the supercell (dark blue spheres). The supercell is repeated periodically to conform periodic boundary conditions (PBC). Shaded spheres indicate atoms which are continuations resulting from the PBC implementation. The indicies $\alpha$, $\beta$ distinguish the component atoms of each molecule.} 
 \label{fig:plane_model}
\end{figure}

We consider hydrogen molecules stacked vertically on a $2D$ square (\emph{x-y}) lattice (cf. Fig.~\ref{fig:plane_model}). This $2D$ molecular crystal is parametrized by the bond-length $R$ and the inter-molecular distance (lattice parameter) $a$.
It must be stressed that even though we consider a \emph{finite} system, we emulate the translational invariance by imposing the periodic boundary conditions (PBC).
The supercell contains four $H_2$ molecules. Let us assign each molecule in the lattice by integers $1,2,3...i,j,k$ etc.
Additionaly, we introduce the indices $\alpha$ and $\beta$ to distinguish the two atoms within the $i-th$ molecule.
Since it is assumed that single--particle states form orthogonal and normalized basis, we have
\begin{align}
\label{eq:orthogonality}
  \bracket{w_{i}^{\mu}(\vec{r})}{w_{j}^{\nu}(\vec{r})} = \delta_{ij}\delta_{\mu\nu},
\end{align}
where $\mu,\nu \in \{ \alpha,\beta \}$.

In this manner, each atom is labelled with the pair $(i,\mu)$ of the indices which also results in the labelling of the microscopic parameters, i.e., $t_{ij} \rightarrow t^{\mu\nu}_{ij}$ and $V_{ijkl} \rightarrow V_{ijkl}^{\mu\nu\tau\rho}$. Effectively, we consider a degenerate two-orbital system.

\begin{figure*}
 {\includegraphics[width=0.35\linewidth]{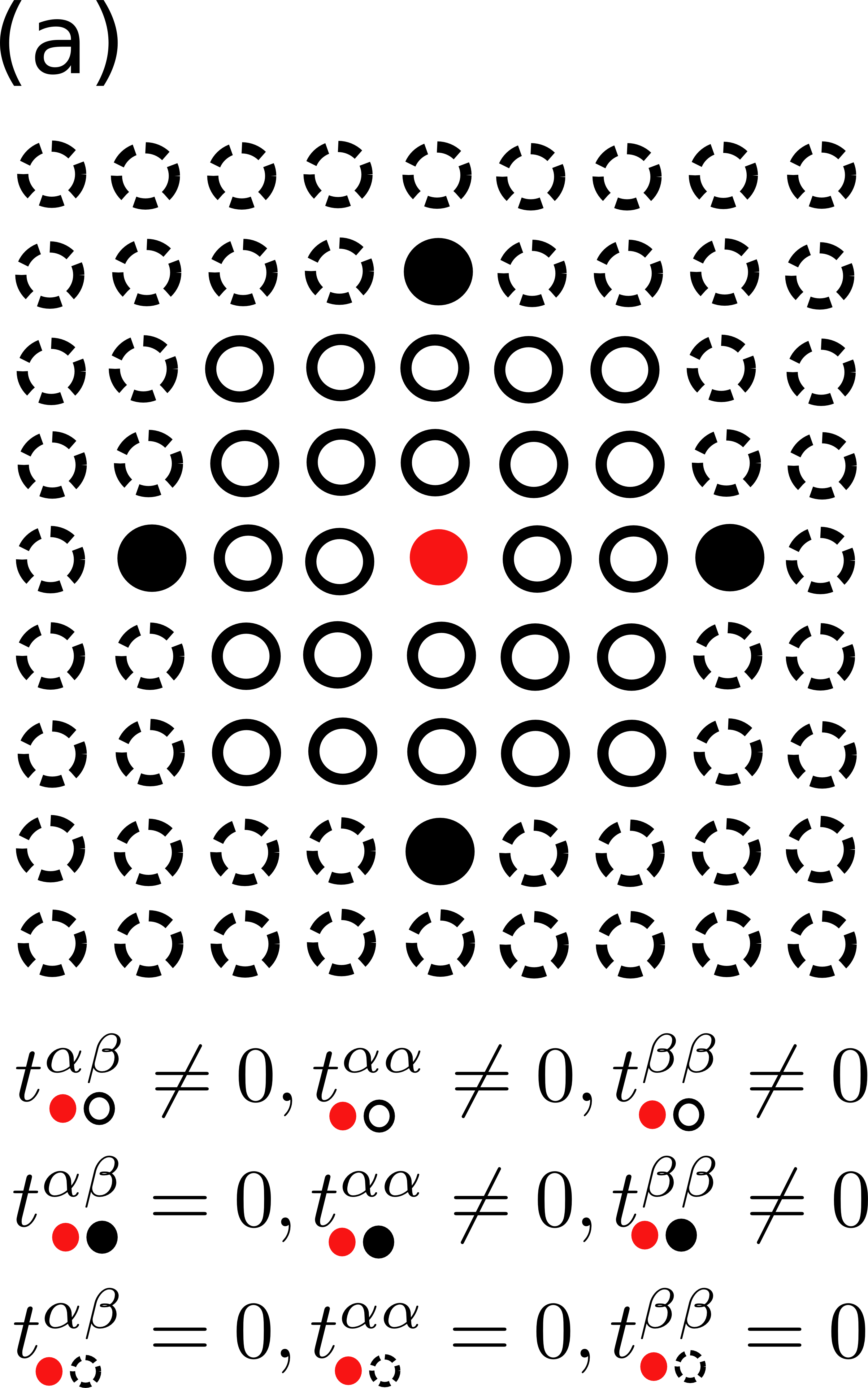}}
 $ $ $ $ $ $ $ $ $ $
 \hfill \includegraphics[width=0.45\linewidth]{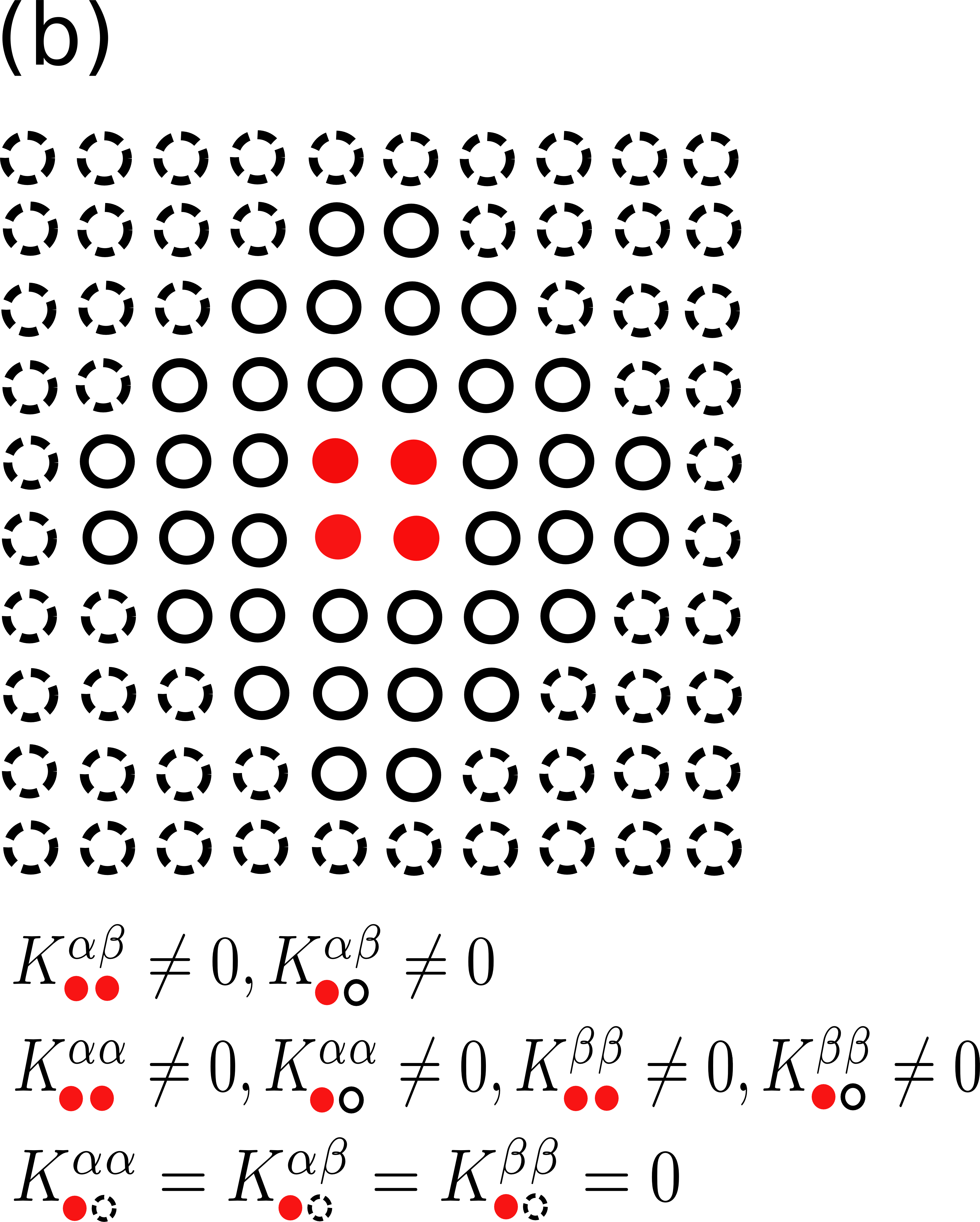}
 \caption{Schematic representation of non-zero hoppings (a), and electron-electron interaction terms (b), taken into account in the Hamiltonian \eqref{eq:hamiltonian_tot}. The corresponding nonzero terms included are listed explicitly and represent the matrix elements between site pairs marked also as the solid or bold circles.}
 \label{fig:hoppings_and_field}
\end{figure*}

 Functions $w_{i}^{\mu}(\vec{r})$ are approximated by means of the \emph{tight--binding} approach, i.e., as a linear combination of $1s$ Slater orbitals which are defined as
\begin{align}
\label{eq:slater}
	\psi_{i}^{\mu} \left(\vec{r}\right) \equiv \sqrt{\frac{\zeta^3}{\pi}} e^{-\zeta\left| \vec{r} - \vec{R}^{\mu}_{i} \right| },
\end{align}
where $\zeta$ becomes single variational parameter to be adjusted in the correlated and $\vec{R}_{i}^{\mu}$ stands for the atomic position, i.e.,
\begin{align}
 \label{eq:wannier_definition}
 w_{i} (\vec{r}) \approx \sum_{j(i)}^{L(i)}\sum_{\mu \in \{\alpha,\beta\}} c_{j\mu} {\psi}^{\mu}_{j} \left(\vec{r} \right),
\end{align}
with the summation related to $j$ extended up to the $13th$ coordination zone, (c.f. Fig.\ref{fig:hoppings_and_field}).The mixing coefficients $c_{j\mu}$ for a given set $\{a,R,\zeta\}$ are to fulfill condition \eqref{eq:orthogonality} within terms of the previously elaborated procedure \cite{Kadzielawa,Biborski}. Both one- and two-electron integrals (Eqs. \eqref{eq:one-body} and \eqref{eq:two-body}, respectively) are also taken into account up to $13$th coordination zone, i.e., extend \textbf{beyond} the supercell and in this sense we include long-range interactions. Note that subscript indices in the hopping and the interaction terms in \eqref{eq:microscopicP} are related to the positions of the atomic centers, i.e., each pair refers to the $|\vec{R}_i - \vec{R}_j|$ distance. We choose the indexing in such a manner that the coordination zone number $z$ for $i =0$ fulfills relation $z = j$. In accordance with our previous investigations \cite{Kadzielawa}, we consider only the two-electron terms with the following coupling constants
\begin{align}
\label{eq:U}
 V_{iiii}^{\mu\mu\mu\mu} \equiv U \ \ \ \text{and} \ \ \ 
 V_{ijij}^{\mu\nu\mu\nu} \equiv K_{ij}^{\mu\nu}, 
\end{align}
with $\mu \neq \nu$ when $i=j$. In effect, taking into account the classical electrostatic interactions between the protons, as well as the interactions within single molecule, the total Hamiltonian describing the system is taken in the form
\begin{align}
 \label{eq:hamiltonian_tot}
 \hat{\mathcal{H}} =& \sum_{i\mu} \epsilon_i^{\mu} \NUM{i}{0} + \sum_{ij\mu\nu\sigma}' t_{ij}^{\mu\nu} \CR{i\mu}{2}\AN{j\nu}{2} \\\notag
 & + U \sum_{i,\mu} \NUM{i\mu}{1} \NUM{i\mu}{-1} + \frac{1}{2} \sum_{ij\mu\nu}' K_{ij}^{\mu\nu} \NUM{i\mu}{0} \NUM{j\nu}{0} \\\notag 
 & + \frac{1}{2}\sum_{ij}\frac{2}{|\vec{R_{i}} -\vec{R_{j}}|},
\end{align}
where $\epsilon_i^{\mu} \equiv t_{ii}^{\mu\mu}$ and $\NUM{i\mu}{0} \equiv \NUM{i\mu}{1} + \NUM{i\mu}{-1} = \CR{i\mu}{1}\AN{i\mu}{1} +\CR{i\mu}{-1}\AN{i\mu}{-1}$.
The primed summations excludes the case of concurrent $i=j$ and $\mu=\nu$. Also, we have neglected here direct exchange-interaction terms and the additional many-site terms, as they are regarded as not essential to the physics of the problem, when considering the threshold of metallicity approached from the molecular side.

\section{\texorpdfstring{Computational method and physical results:\\
From 2D molecular crystal to quasiatomic metallic bilayer}{Computational method and physical results: From 2D molecular crystal to quasiatomic metallic bilayer}}
\label{sec:results}
\subsection{The enthalpy and the pressure definition in two dimensions}
The $2D$ system is studied here under action of an external side force (effective pressure). However, in such a two-dimensional situation the \emph{pressure} has to be redefined. Namely, an external homogeneous force is exerted on the $2D$ crystal in the planar (\emph{x-y}) directions. Therefore, this situation is a $2D$ analog of the action of hydrostatic pressure onto a three-dimensional system. The elementary volume of $2D$ crystal is simply 
\begin{align}
 \label{eq:volume}
 v_{2D} &\equiv a^2,
 \end{align}
and thus, the pressure in the present case is
\begin{align}
 \label{eq:pressure}
 p \equiv p_{2D} &\equiv \frac{f}{a},
 \end{align}

where $f$ is the force per \emph{"unit cell"} exerted homogeneously on the system in the planar directions. By taking this definition of pressure we have the usual definition of work part of the internal energy (or the enthalpy) as $f a = p v_{2D}$. Note that an \emph{infinite} nature of the system is considered here preserved by means of applying PBC. Finally, a proper function of state, which in this case is $2D$ enthalpy per molecule can be defined as
\begin{align}
 \label{eq:enthalpy}
 h &\equiv \frac{E(a,R)}{N} + p_{2D} v_{2D},
 \end{align}
where $E(a,R)$ is the ground state energy for given structural parameters $a$ and $R$ (c.f Fig.~\ref{fig:plane_model}).
We scan the space $(a,R)$ of the parameters to obtain the energy landscape of $E(a,R)$. Note that the meaning of $f$ arises from the notion that the enthalpy should be defined as an extensive function of the system volume $v_{2D}N$, where $N$ is the number of molecules in the system. Also, as an outcome of our approach, we obtain evolution of the system as a function of the applied force as the only independent variable, i.e., $E(a,R) \equiv E(a(f),R(f))$. In this manner, the theory is fully microscopic, as all the microscopic parameters of the Hamiltonian \eqref{eq:hamiltonian_tot}, as well as $a$ and $R$, are determined explicitly, within our EDABI procedure.

\subsection{Computational details}
The whole procedure is composed of the three stages: (\emph{i}) selection and orthogonalization of the starting trial basis $\big\{{w_{i}^{\mu}(\vec{r})}\big\}$, (\emph{ii}) calculation of integrals $t_{ij}^{\mu\nu}$ and $K_{ij}^{\mu\nu}$, and (\emph{iii}) diagonalization of Hamiltonian matrix and concomitant minimization of the ground state energy with respect to $\{\lambda\}$. 

The orthogonal single particle basis is obtained in (\emph{i}) in terms of the numerical solution of the bi-linear set of equations \eqref{eq:orthogonality} with the desired accuracy ($10^{-6}$ in our case is assumed as sufficient).
Step (\emph{ii}) is also performed numerically by means of the previously elaborated method \cite{Biborski}. Each of the Slater $1s$ orbitals, which are the \emph{building blocks} of $\{w_{i}^{\mu}(\vec{r})\}$ functions (see Eq. ~\ref{eq:wannier_definition}), are approximated by three Gaussian functions what simplifies the calculation of the two-electron integrals composing $\{V_{ijkl}^{\mu\nu}\}$ \cite{Kadzielawa,Biborski}. Note also that according to the spatial \emph{cutoff} assumed for the repulsive Coulomb interactions, there are $23 + 1 =24$ (intersite plus one intrasite, respectively) $K_{ij}^{\mu\nu}$ integrals to be computed, carried out each time when the variational parameter $\zeta$ is updated during the minimization procedure. This stage is the most time consuming in the whole procedure. The step (\emph{iii}), i.e., the Hamiltonian matrix diagonalization, is performed for the moderately sized matrix ($12870 \times 12870$), and results from the assumed model, i.e., that with the half filling for the $8$-site system. The periodic booundary conditions (PBC) are imposed in the standard manner by means of inclusion of \emph{up-to-cutoff} terms in the Hamiltonian matrix (cf. Fig.~\ref{fig:hoppings_and_field}) which is diagonalized subsequently with the help the Lanczos algorithm. The diagonalization of \eqref{eq:hamiltonian_tot} results thus in obtaining the trial value of the trial ground state energy $E_G(\zeta)$. The latter is minimized with respect to $\zeta$ by means of numerical procedure devoted for a single variable function numerical scheme (e.g., \emph{Brent}, as in this case or \emph{golden section search}), implemented within the \emph{Gnu Scientific Library} (GSL) used by us in this context. The typical numerical accuracy of the energy evaluation is $10^{-4}$ Ry. As the phase transition to the quasiatomic phase is of the first-order nature, such accuracy is sufficient as we can trace the evolution of the involved enthalpies in a systematic manner, as a function of applied pressure.

\subsection{Discontinuous \texorpdfstring{$H_2 \rightarrow 2H$}{H2 -> 2H} transition and its overall characteristics}
\begin{figure}
 \includegraphics[width=\linewidth]{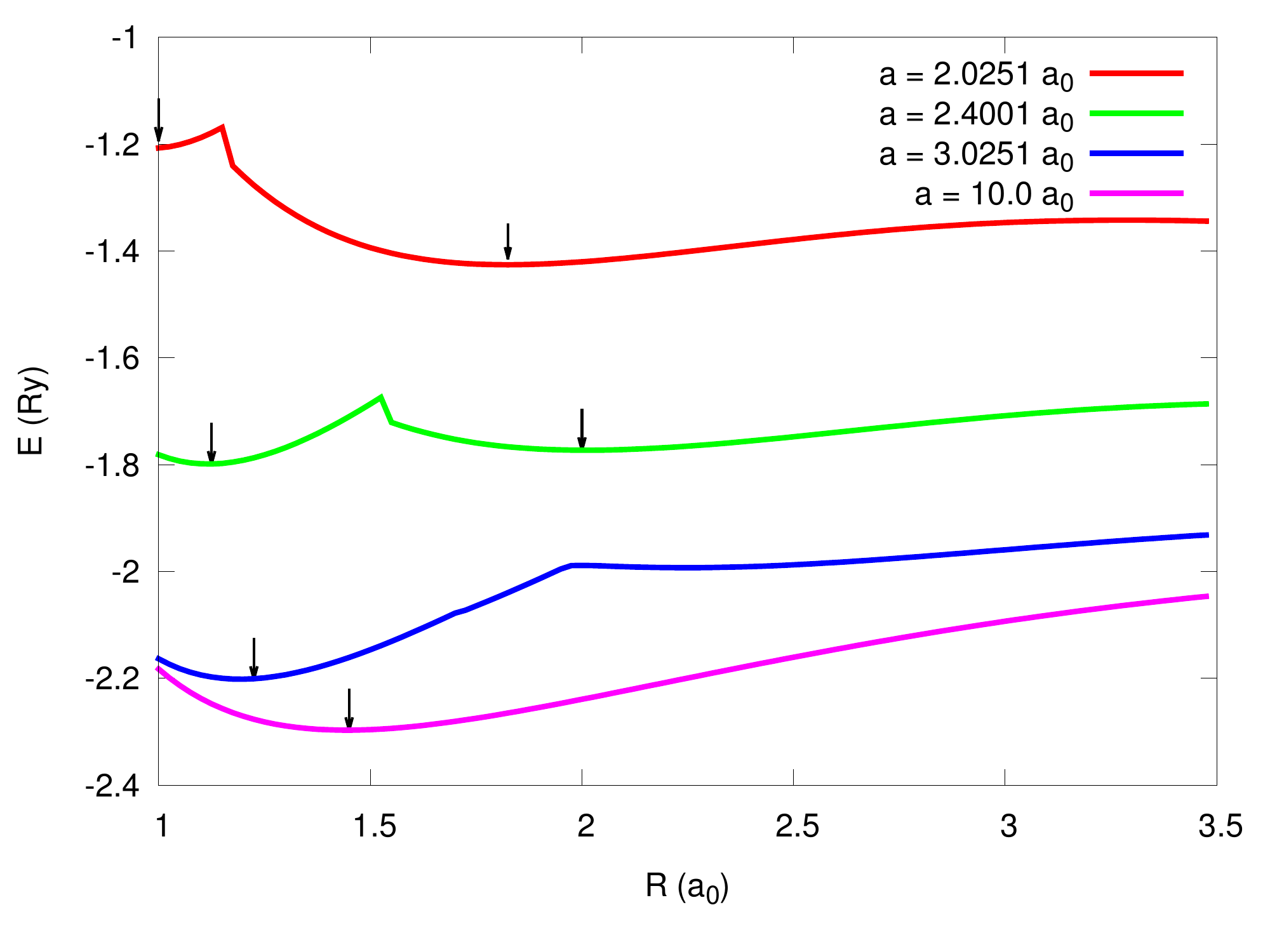}
 \caption{Ground-state energy per molecule as a function of the bond length (intramolecular distance) $R$ for four selected values of the lattice parameter $a$. The minima are marked by the vertical arrows}
 \label{fig:2d_energy}
\end{figure}

\begin{figure}
 \includegraphics[width=\linewidth]{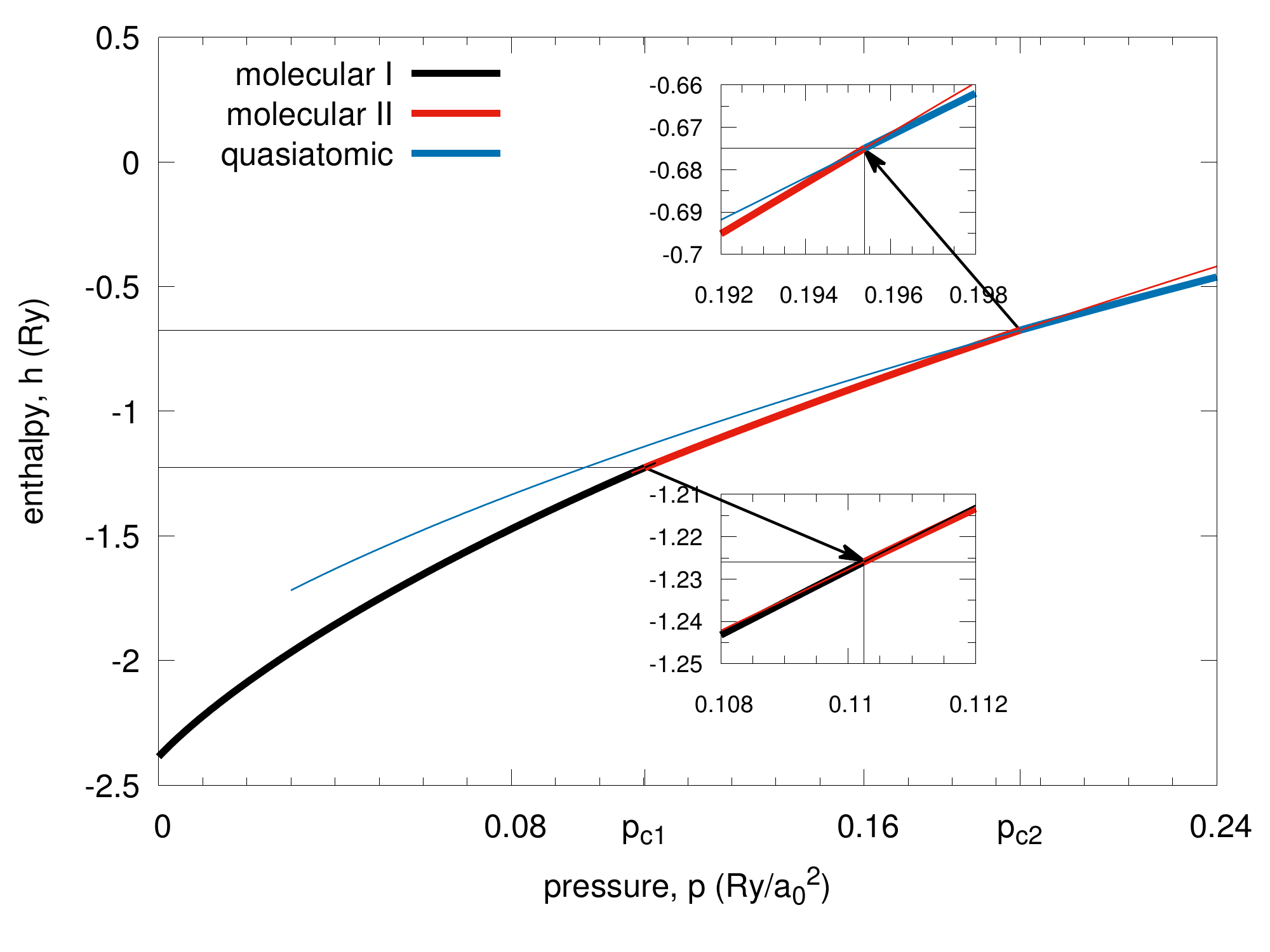}
 \caption{The enthalpy (per molecule) versus pressure $p$. At lower pressure, two molecular phases are stable; the transition to the \emph{quasiatomic} phase
 occurs at $p_{c2} \sim 0.1954 Ry/a_{0}^2$, as marked. $E_B(p=0)=-2.3858 Ry$, $R_{eff}(p=0) = 1.4031 a_0$, $a(p=0) = 4.3371 a_0$. Thin lines extrapolate the enthalpies of the particular phases beyond the regime of their stability. Insets show some detail on the transitions. For details see main text.}
 \label{fig:2d_enthalpy}
\end{figure}

We start our discussion with remark that the solid hydrogen dissociation from molecular to the quasiatomic state, and associated with it metallization, represents one of the the fundamental transitions in Nature, as it involves one of the simplest condensed systems in which the electronic correlations play a decisive role, as we discuss next. 
In Fig.~\ref{fig:2d_energy} we present exemplary results for the ground-state energy versus the bond length $R$ for the four selected values of of the lattice parameter $a$. With the decreasing $a$, the molecular bond length evolves from the value $R \ll a$ at ambient pressure to that close $a$. Such a changeover speaks directly about the transition from molecular to quasiatomic configuration. The detailed character of the transformation is shown in Fig.~\ref{fig:2d_enthalpy},
where we have displayed the enthalpies of two molecular ($R \ll a$) phases and the atomic one ($R \sim a$) as a function of applied pressure. Two discontinuous (first-order) phase transitions are seen at the critical pressures $p_{c1} \sim 0.1102 Ry/a_{0}^2$ and $p_{c2} \sim 0.1954 Ry/a_{0}^2$, respectively, where $a_0$ is the Bohr radius. Note that at $p = 0$ the equilibrium values of the binding energy and the bond length are $E_B = -2.3858 Ry$ and $R = 1.4031 a_0 = 0.7425 \text{\normalfont\AA}$, respectively. These values can be compared with those for $H_2$ molecule: $E_B = -2.295Ry$ and $R=0.74144\text{\normalfont\AA}$ \cite{Kolos}. So the solid molecular bilayer is stable against the dissociation into individual molecules and the bond length in the former case is larger  by $0.14\%$. This result provides \emph{a crucial test of our method} reliability when applied to the multimolecular systems. Obviously, the values of $E_B$ at $p=0$ prove only that the solid molecular phase is stable for $p < p_{c2}$ from the electronic point of view, as we have not included as yet the zero-point motion. Those will be estimated later. The application of pressure will help additionally to stabilize it.

\begin{figure}
 \includegraphics[width=\linewidth]{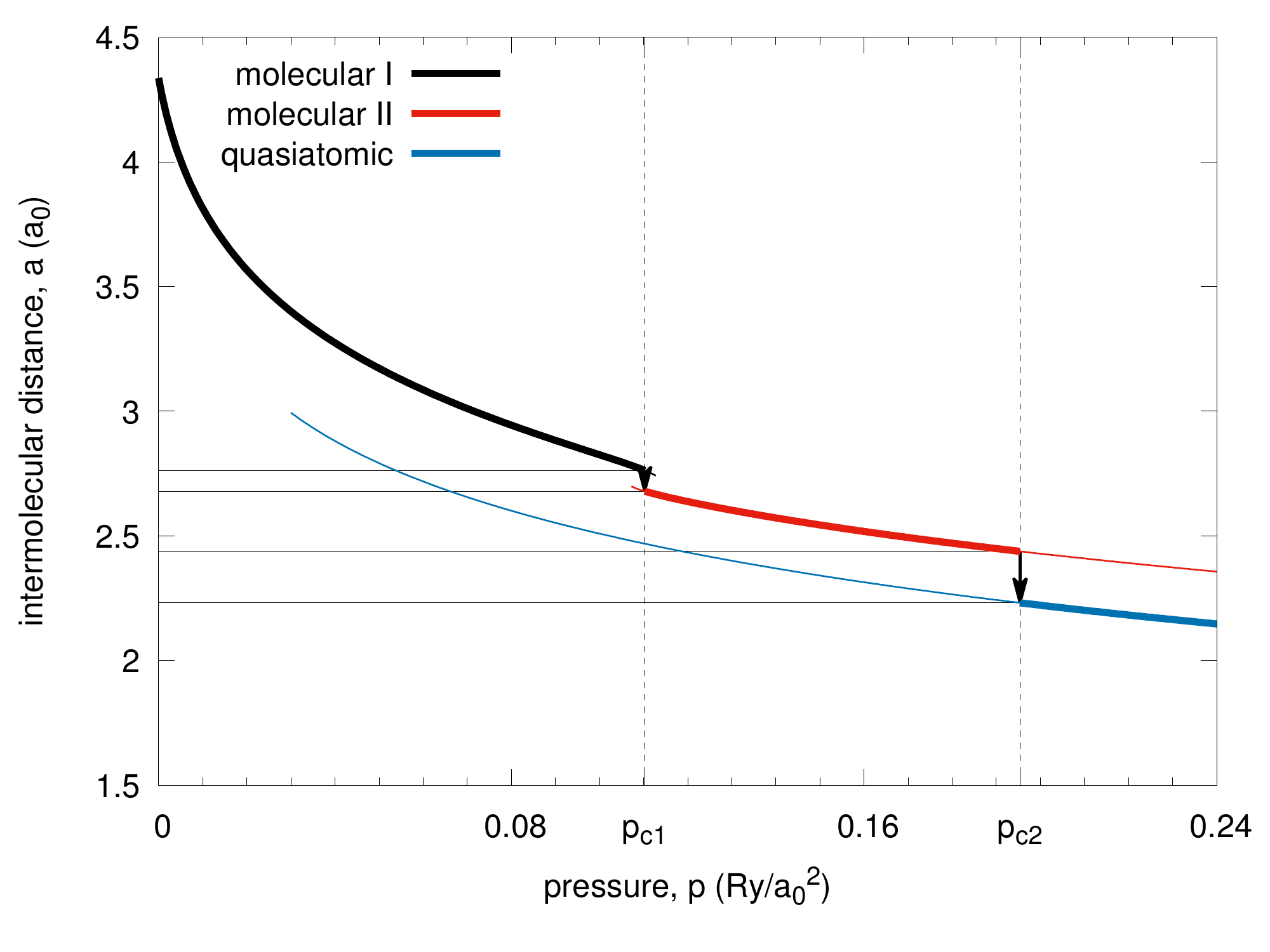}
 \caption{Intermolecular distance (lattice parameter) $a$ for $2D$ bilayer crystal as a function of pressure $p$. The transitions are clearly of discontinuous (first-order) nature at temperature $T=0$. Note a spectacular decrease of lattice parameter by $8.47\%$ (corresponding to $16.22 \%$ volume decrease) at the transition ($p_{c2}$) from molecular to quasiatomic phase. The thin lines denote the lattice parameter of the phases in the regime, where they are not of the lowest enthalpy. The arrows marks the jump of the intermolecular distance at the transitions with the increasing pressure.}
 \label{fig:2d_a}
\end{figure}

\begin{figure}
 \includegraphics[width=\linewidth]{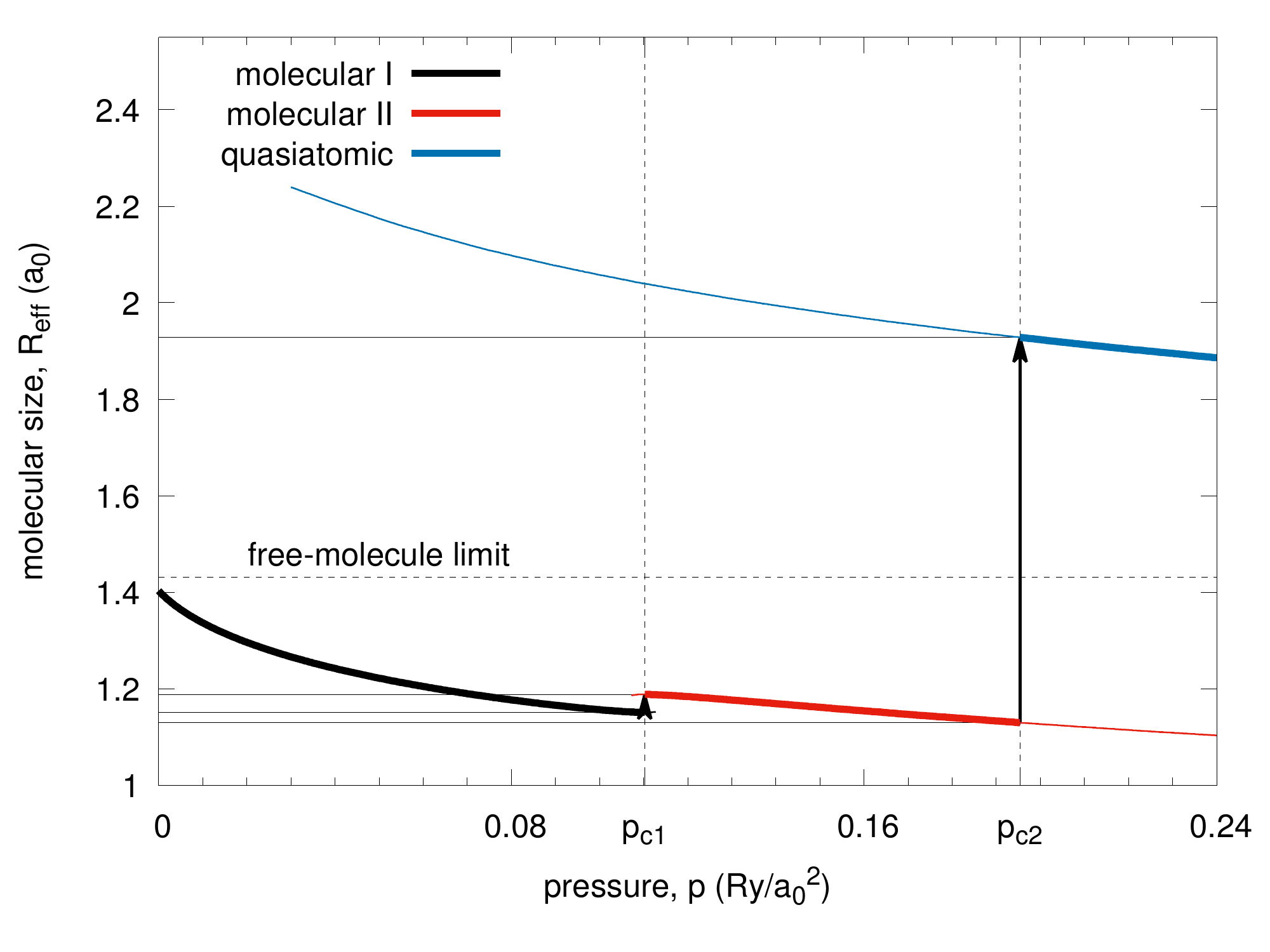}
 \caption{ Intramolecular distance (bond length) $R$ as a function of pressure $p$. An abrupt change by $70.69 \%$ at the transition from molecular to quasiatomic state (at $p_{c2}$) is clearly visible. The spectacular increase of the optimized bond length $R=R_{eff}$ at $p_{c2}$ is taking place towards quasiatomicity (cf. Fig. ~\ref{fig:2d_a}). Only a small difference between $R_{eff}$ in both of the molecular phases ($3.21 \%$ -- close to $p_{c1}$) is observed. The arrows mark the interatomic distance jump at $p_{c1}$ and $p_{c2}$ when increasing $p$.}
 \label{fig:2d_R}
\end{figure}

\begin{figure}
 \includegraphics[width=\linewidth]{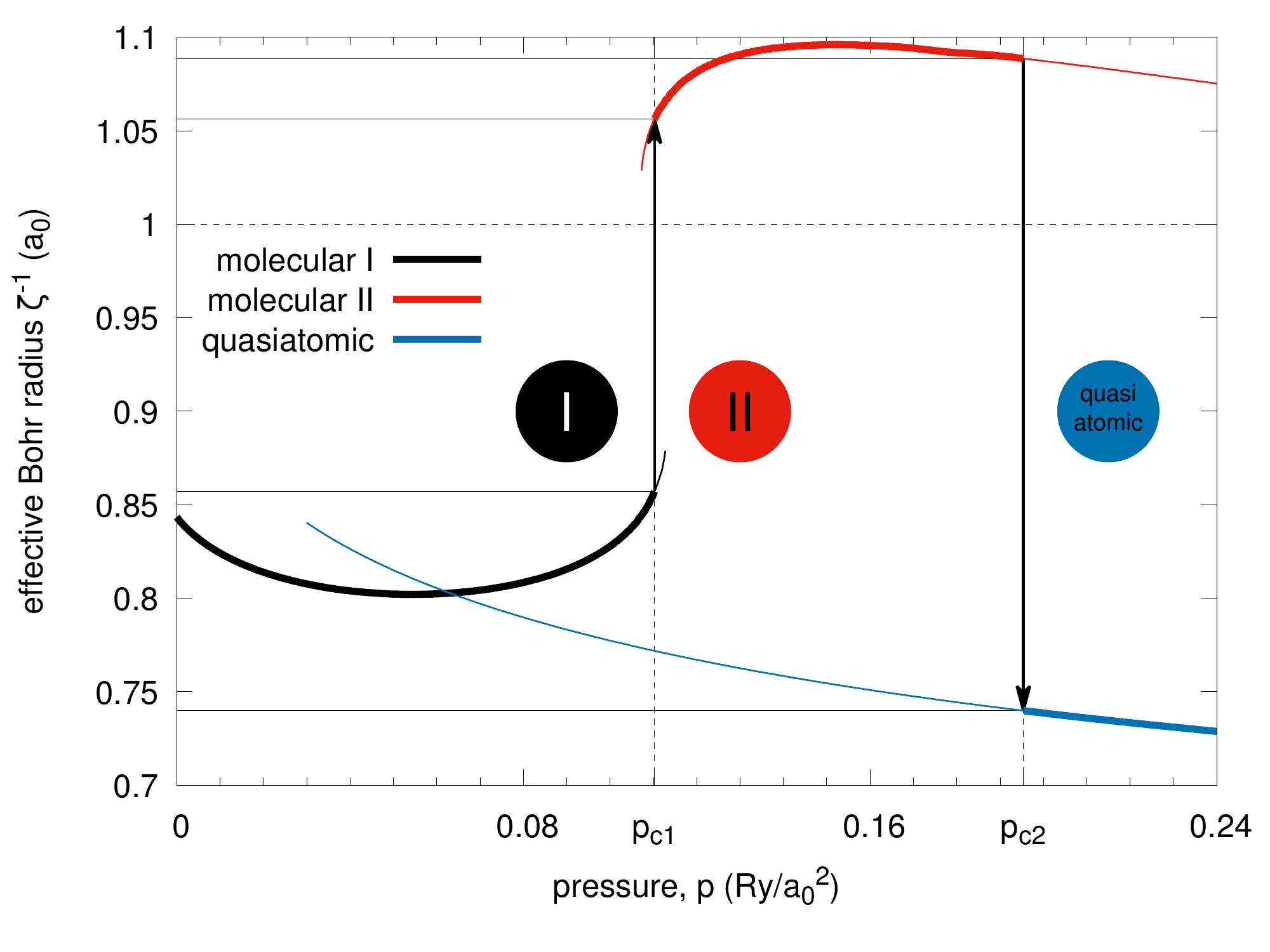}
 \caption{ The effective Bohr radius $1/\zeta$ of the renormalized Slater orbitals composing the Wannier functions for $2D$ system, as a function of pressure $p$. The atomic function size changes by $27.02 \%$ at the transition to the molecular phase II and by $29.97 \%$ at the transition to the quasiatomic (metallic) state. The arrows mark the Slater-orbital size jumps when increasing $p$.}
 \label{fig:2d_zeta}
\end{figure}

In Fig.~\ref{fig:2d_a} we plot the equilibrium lattice parameter (in units of $a_0 \approx 0.53 \text{\normalfont\AA}$) versus pressure and observe a discontinuous lattice contractions for both the transitions by about $3\%$ and $9\%$ at the pressures $p_{c1}$ and $p_{c2}$ respectively. In an analogous manner, the bond length vs pressure jumps from the equilibrium value $R_{eff} \ll a$ to $R_{eff} \sim a$ at the critical pressure $p_{c2}$, as shown in Fig.~\ref{fig:2d_R}. Hence the transitions are strongly discontinuous between the each of the two pair of three stable phases. The phase diagram for the scanned space of $(a,R)$ is composed of three phases. Those referring to $p\leq p_{c1}$ and $p_{c1} \geq p \leq p_{c2}$ we recognize as both being of a \emph{molecular} kind and label them them as phases I and II, respectively, while the phase referring to $p \geq p_{c2}$ is the \emph{quasiatomic} one. This distinction may seem at this stage as somewhat arbitrary and is legitimate only by making observation that the ratio $a/R \geq2$ for stable phase referring to $p\leq p_{c2}$ and $a/R \approx 1$ for $p\geq p_{c2}$. However, more convincing argument which originates from the diversity of electronic properties for both of the two groups of phases, is provided in the next subsection. As a supplementary information we have plotted in Fig.~\ref{fig:2d_zeta} the inverse Bohr radius $\zeta^{-1}$ vs $p$ for the Slater functions composing the Wannier functions. The jumps take place by $\sim 27\%$ at $p_{c1}$ and by $\sim 30\%$ at $p_{c2}$, so the wave-functions site is  strongly altered at both the transitions. Note that $\zeta^{-1}$ value in the $H_{2}$ phase is close to that for the hydrogen atom (within $\sim 3\%$) even though the actual value in the quasiatomic solid phase is only about $75\%$ of the single-atom value. This last results is certainly counterintuitive. Interelectronic correlations, induced by the interatomic repulsive interactions, reduce the effective Bohr radius by over $17\%$ in the molecular phase II.

\subsection{Principal electronic characteristics of the Mott-Hubbard \texorpdfstring{$H_2 \rightarrow 2H$}{H2 -> 2H} transition}

\begin{figure}
 \includegraphics[width=\linewidth]{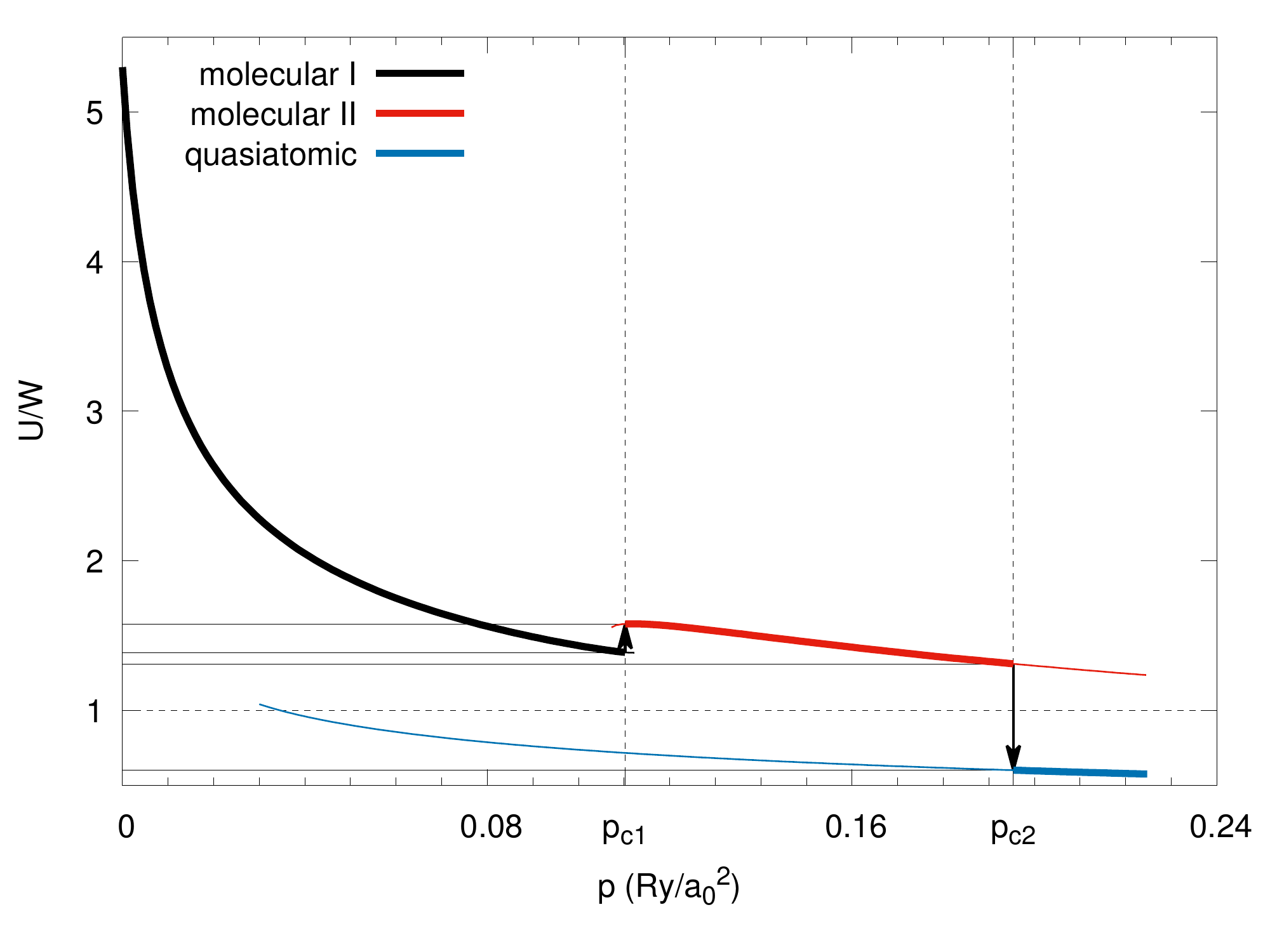}
 \caption{The ratio between the intraatomic (Hubbard) repulsion amplitude $U$ and the lower-bandwidth $W$ in the correlated state, as a function of $p$. Both quantities are calculated for the renormalized orbitals composing the Wannier functions. At the critical pressures the ratio jumps:
 from the value $1.3880$ to $1.5778$ at $p_{c1}$ (at the transition between the two molecular phases), and from $1.3112$ to $0.6000$ at $p_{c2}$, i.e., at the transition to the quasiatomic phase. The latter defines the Mott-Hubbard-type transition to a moderately correlated state. Close to the transition, even in the molecular phases the value of the bare bandwidth $W$ is not decisively smaller than $U$. The arrows mark the jumps when increasing the pressure.
 }
 \label{fig:2d_UW}
\end{figure}

For the sake of completeness, we list in Table~\ref{tab:wf_coeffs} principal parameters of the three states calculated at the critical pressures. Particularly interesting are $t^{\alpha\beta}_{00}$ and $t^{\alpha\alpha}_{01}$, the intra- and inter-molecular hopping integrals, since they change remarkably at the transition. The same concerns the values of the Hubbard gap $U\! -\!W$ (with the bare bandwidth $W$ calculated in Appendix A) and the $U/W$ ratio (c.f. Figs.~\ref{fig:2d_W} and ~\ref{fig:2d_UW}, respectively). The last characteristic is particularly important since at the transition at $p_{c2}$ it jumps from $U/W = 1.3112$ ($> 1$), in the molecular state to the value $0.6000$ ($ < 1$) and represents a typical trend for the Mott-Hubbard transition, albeit this time from an originally \emph{ diamagnetic molecular insulator to a paramagnetic metal}. The negative value of the Hubbard gap means that the two lowest bands overlap appreciably and therefore the system can be regarded as metallic.Also, there is a principal difference between the present approach and the the canonical treatments \cite{Gebhard,Fazekas,Fulde} of the Hubbard model, as here the value of the bandwidth changes at the transition, and in effect, the $U/W$ ratio is, not as one would have in all the parametrized-model considerations \cite{Hubbard2,Brinkman,Spalek2}, changing in a continuous manner. Also, a relatively large value of the intersite Coulomb interactions may mean that either the spin (SDW)- or the charge (CDW)-density-wave states become a stable phase on the quasiatomic side, at least in the low-temperature range. This topic should be analyzed separately, as it is more complicated than the present analysis. Such an analysis would allow for differentiating in detail between the present transition from the diamagnetic insulator and the canonical Mott-Hubbard transition which takes place from an antiferromagnetic (Mott) insulator to either SDW or a paramagnetic correlated metal. Also, as said above, the Mott-Hubbard transition is analyzed customarily as a function of $U/W$ ratio changing continuously \cite{Hubbard2,Brinkman,Spalek2,Vollhardt}. As our results show explicitly this is not the case, when the renormalization (readjustment) of the orbitals is taken into account in the correlated state. In this respect, our approach is fully microscopic (parameter free).

\begin{figure}
 \includegraphics[width=\linewidth]{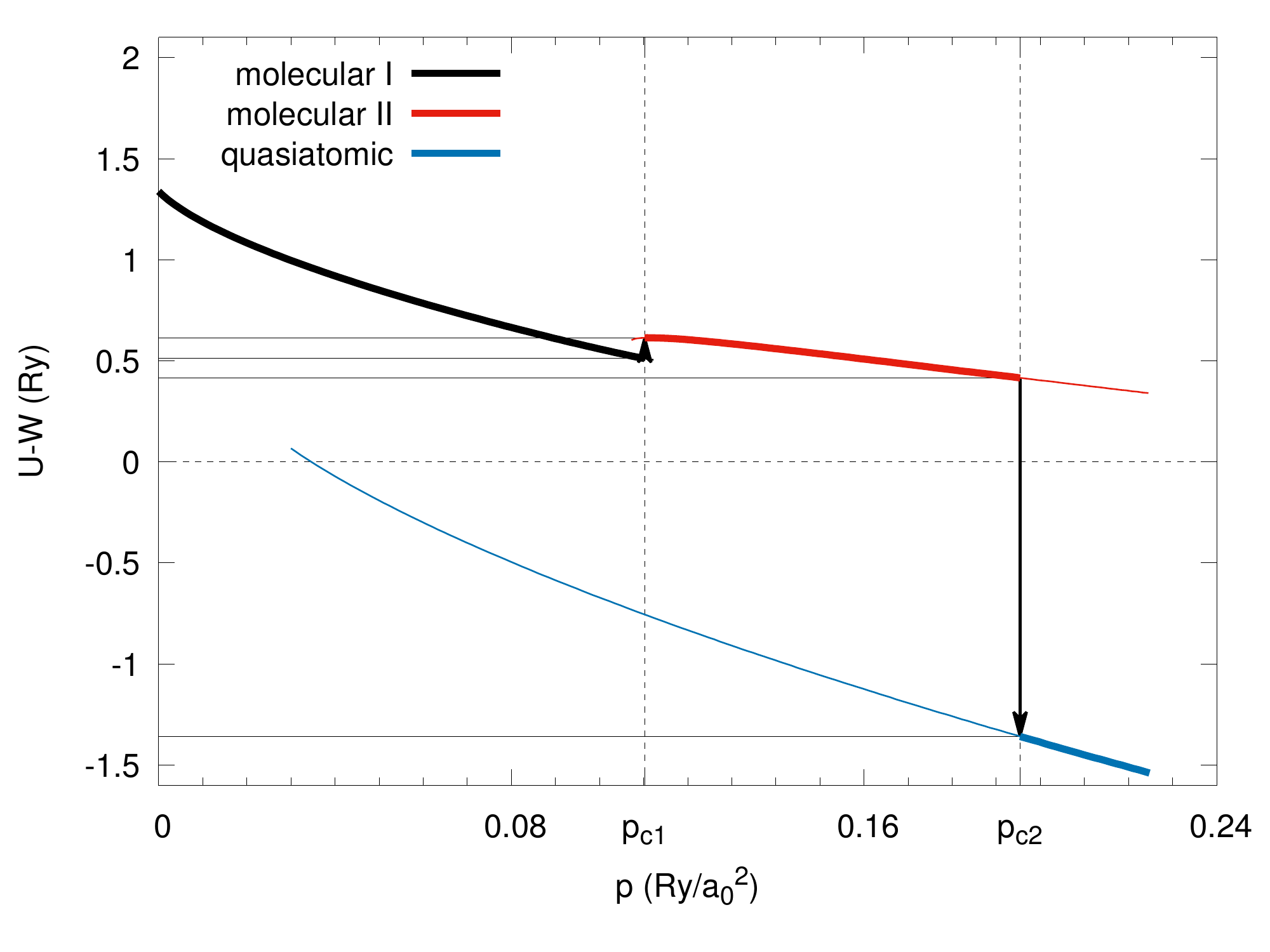}
 \caption{Estimate of the Hubbard gap, $U-W$, with the bare bandwidth $W$ computed for the single-electron part of Hamiltonian \eqref{eq:hamiltonian_tot} as a function of $p$ in the molecular and quasiatomic correlated states. The bandwidth changes radically at $p_{c2}$. The negative gap value means that the two bands overlap and hence the system is in metallic state (for a detailed discussion see main text). The arrows mark the sequence of jumps with the increasing pressure.}
 \label{fig:2d_W}
\end{figure}

\begin{figure}
   \includegraphics[width=0.3\linewidth]{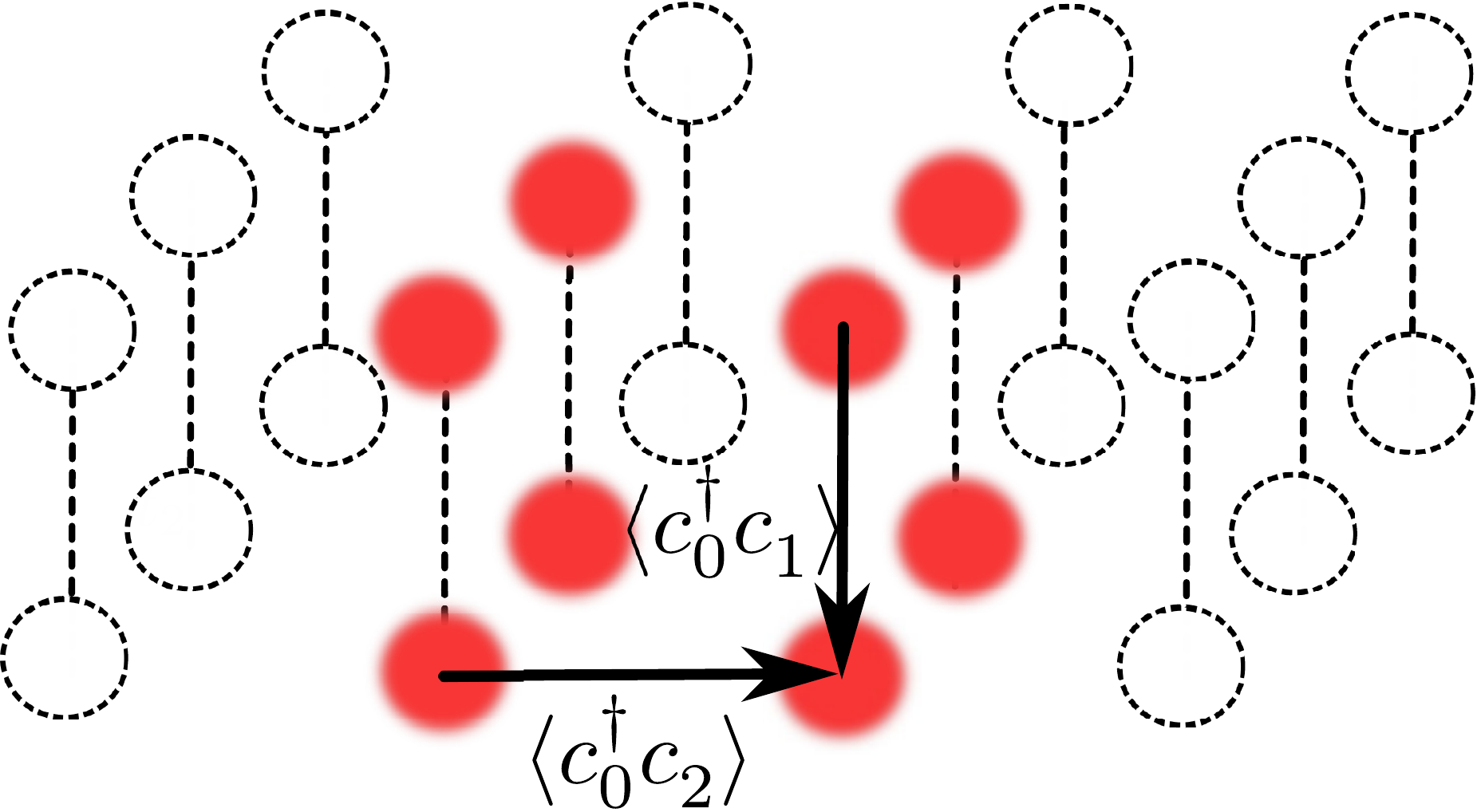}\
   \includegraphics[width=0.8\linewidth]{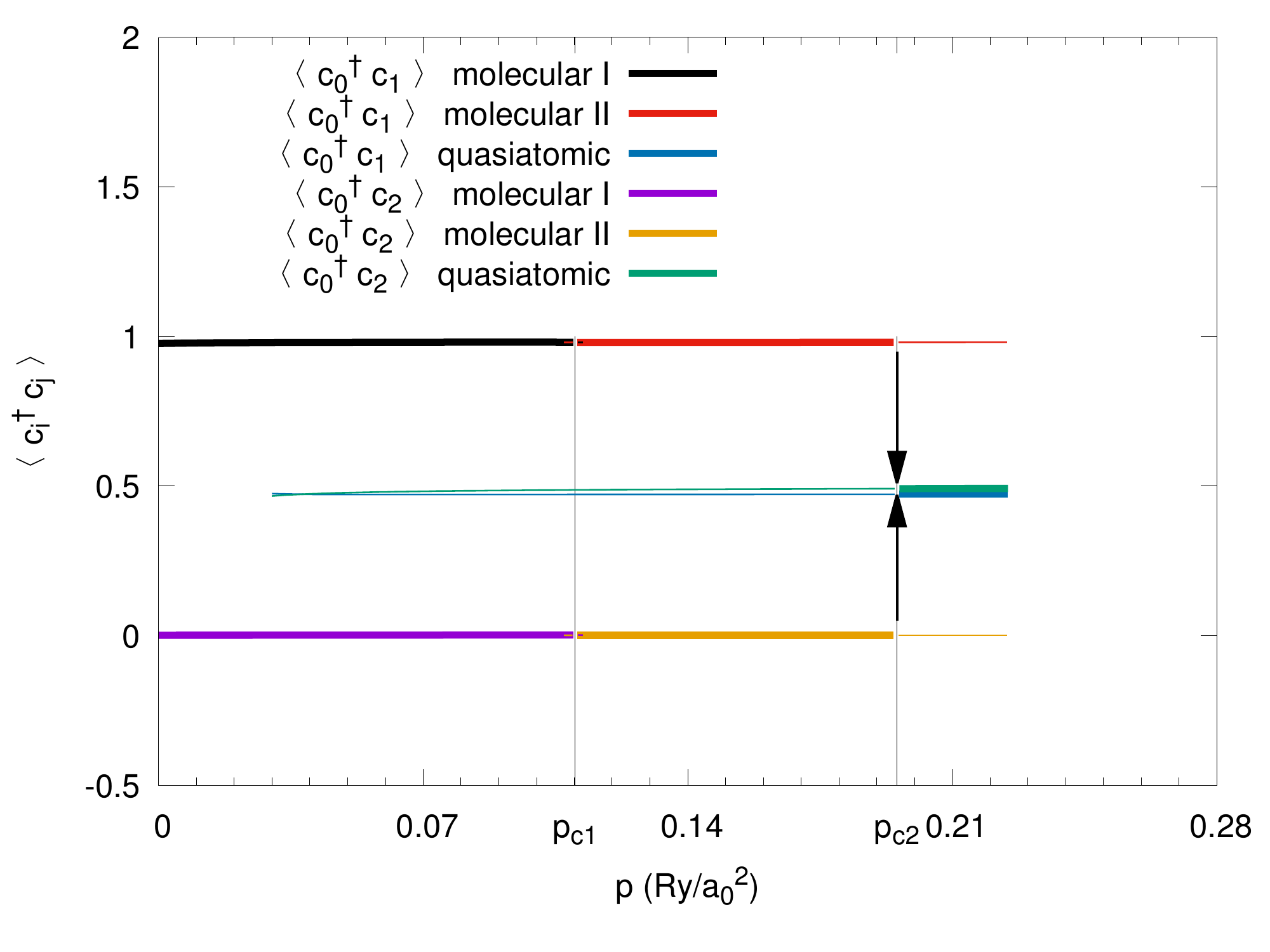}
 \caption{Principal hopping correlation functions $\average{\CR{i}{0}\AN{j}{0}}$ versus pressure $p$. $\average{\CR{0}{0}\AN{1}{0}}$ corresponds to the intramolecular hopping, $\average{\CR{0}{0}\AN{2}{0}}$ to the intermolecular one; the notation is explained in the upper part of the Figure. Note that whereas for the molecular crystal the dominant hopping is $\average{\CR{0}{0}\AN{1}{0}}=1$ and the remaining one is almost equal to zero, for the quasiatomic phase the presented correlation functions in the metallic state are almost equal to those for free-electrons, i.e., $\approx 1/2$. Such a behavior provides us with a clear sign of both quasiatomic nature and metallic character of the highest-pressure state, as the renormalized hoppings are practically the same and equal to $\tfrac{1}{2}$.
 }
 \label{fig:2d_avgs}
\end{figure}

\begin{figure}
 \includegraphics[width=.9\linewidth]{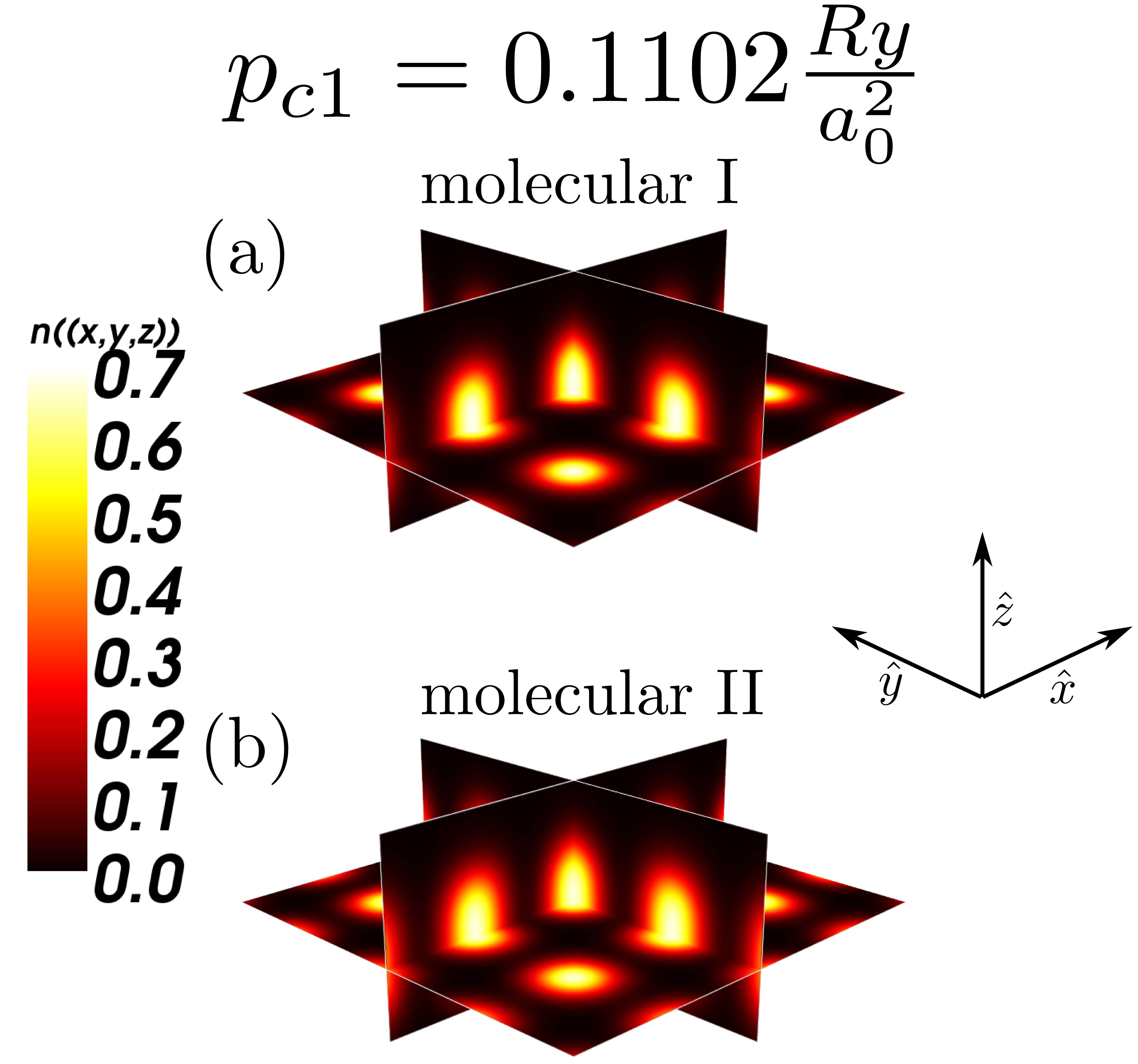}
 \caption{The electronic density $n(\vec{r})$ in 3d near the molecular I (a) ~\textrightarrow~molecular II (b) transition at $p_{c1} = 0.1102 Ry a_0^{-2}$. The ellipsoidal character of density is a signature of $H_2$ molecular states with the symmetric character (with respect to the molecule center of mass) of its spatial distribution.
 }
 \label{fig:2d_densT1}
\end{figure}
\begin{figure}
 \includegraphics[width=.9\linewidth]{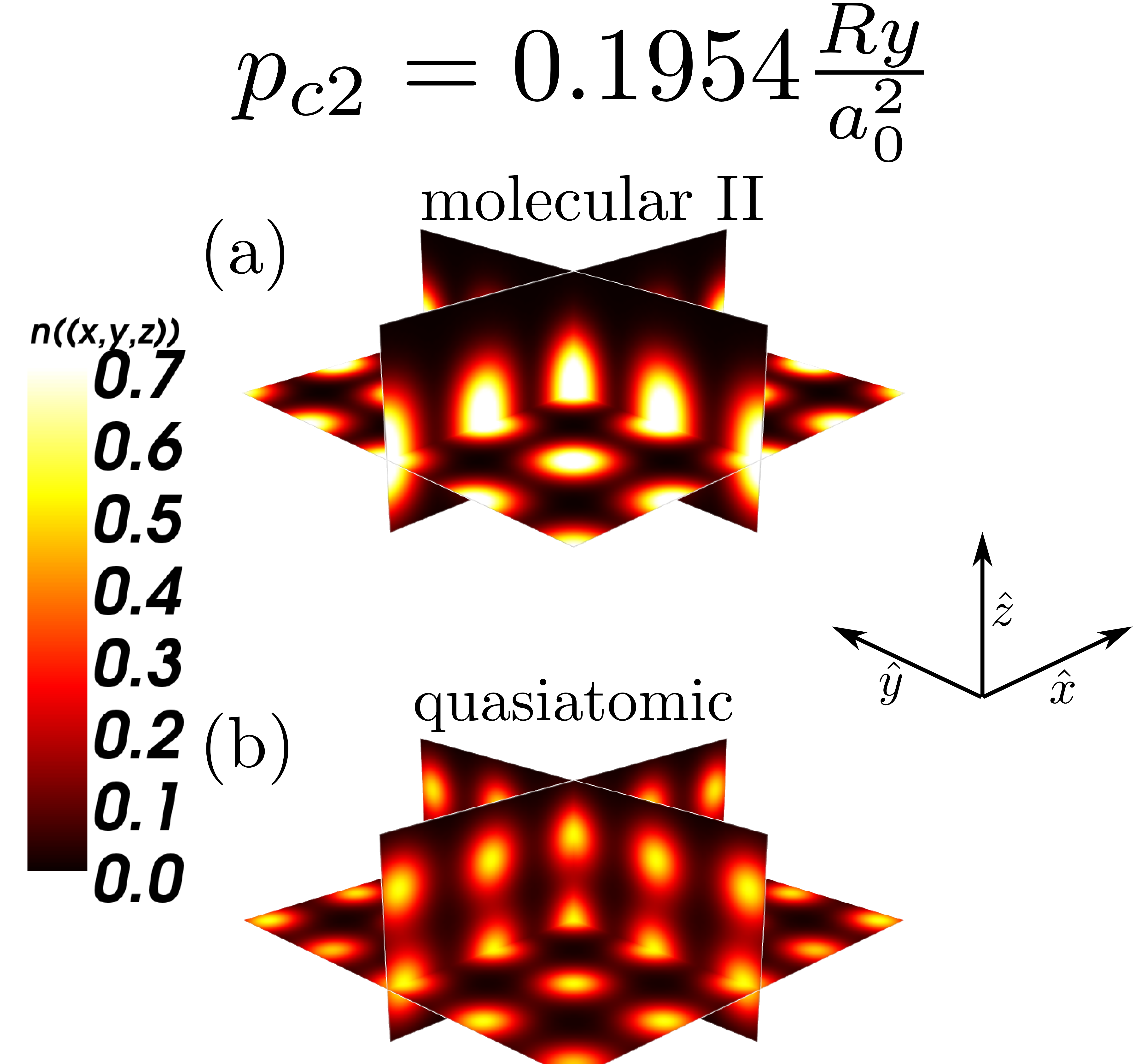}
 \caption{The electronic density $n(\vec{r})$ in 3d near the molecular II (a) ~\textrightarrow~quasiatomic (b) transition ($p_{c2}= 0.1954 Ry a_0^{-2}$). Note a clear changeover from the molecular ellipsoidal (top) to the quasiatomic (spherical) configuration shape of the density, characteristic for symmetric-in-space molecular states and quasiatomic nature of the single-particle states, respectively. Also, the electronic-density profiles illustrate directly the character of the Mott-Hubbard transition at $p=p_{c2}$.
 }
 \label{fig:2d_densT2}
\end{figure}

The transition can be elaborated further by calculating directly the intramolecular (\textav{\CR{0}{0}\AN{1}{0}}) and the intermolecular hopping correlation functions, both displayed in Fig.~\ref{fig:2d_avgs}. Note that the value of correlation function $\textav{\CR{i}{0}\AN{j}{0}} \equiv \sum_\sigma \textav{\CR{i}{2}\AN{j}{2}}$ reaches the value $\tfrac{1}{2}$ in the quasiatomic phase which we identify with the system metallicity. This is because this value reaches an amazing value $n(1-\tfrac{n}{2})= \tfrac{1}{2}$ for $n=1$ electrons per atom, characteristic of the uncorrelated lattice fermionic gas \cite{Spalek3}. On the contrary, the value of \textav{\CR{0}{0}\AN{2}{0}} in the molecular phases is close to zero, whereas $\langle\CR{0}{0}\AN{1}{0}\rangle \approx 1$ then, both characteristic of a molecular insulator. It is amazing that so spectacular switching from an almost ideal insulator to an almost ideal fermionic gas takes place in this situation. The situation described here is in accord with an old argument of Mott \cite{Mott2} that switching to a metallic state can take place only in a discontinuous manner as a creation of a small number of carriers in a nominally insulating state would largely increase the system energy due to the lack of screening of the long-range repulsive Coulomb interaction between them. Here, this argument is fully qualified and includes also the Hubbard argument \cite{Hubbard} in the same manner. In effect, the solid hydrogen may be indeed regarded as the model example of the transition from a correlated, albeit diamagnetic insulator to a moderately/weakly correlated paramagnetic metal, if we only account properly for its original molecular $H_2$ structure in a solid at ambient pressure, and subsequently the renormalization of both the molecular and the atomic (Slater) orbitals by the interelectronic correlations. The values of the lattice and microscopic parameters at the transitions are listed in Table~\ref{tab:wf_coeffs}.

\begin{table*}[t!]
\caption{Values of the principal parameters at both the transition pressures and on both side of those discntinuous transitions. For explanation of notation see Fig. ~\ref{fig:hoppings_and_field} and main text. The numerical accuracy is at the level of the last digit.
}
\label{tab:wf_coeffs}
\resizebox{\textwidth}{!}{
 \begin{tabular}{c||r|r|r|r|r|r|r|r|r|} & $p (Ry/a_0^2)$ & $a (a_0)$ & $R_{eff} (a_0)$ & $\zeta (a_0^{-1})$ & $U (Ry)$ & $K_{00}^{\alpha\beta} (Ry)$ & $K_{01}^{\alpha\alpha} (Ry)$ & $t_{00}^{\alpha\beta} (Ry)$ & $t_{01}^{\alpha\alpha} (Ry)$\\\hline 
molecular I & $0.1102$ & 2.7626 & 1.1511 & 1.1667 & 1.8268 & 1.0725 & 0.7173 & -1.1985 & -0.1933 \\\hline
molecular II & $0.1102$ & 2.6791 & 1.1881 & 0.9466 & 1.6751 & 0.9847 & 0.7289 & -1.1177 & -0.1422 \\\hline\hline
molecular II& $0.1954$ & 2.4378 & 1.1296 & 0.9186 & 1.7486 & 1.0244 & 0.7945 & -1.2456 & -0.1596 \\\hline
quasiatomic& $0.1954$ & 2.2313 & 1.9281 & 1.3516 & 2.0392 & 0.9380 & 0.8760 & -0.7660 & -0.3884
 
\end{tabular}
}
\end{table*}

\subsection{Electron density evolution and renormalized single-particle band characteristics in the correlated state}

To complete our picture of the metallization we have also determined the electron densities $n(\vec{r}) \equiv \textav{\hat{\Psi}^{\dagger}(\vec{r})\hat{\Psi}^{}(\vec{r})}$ in the many-particle states; those are displayed in Figs.~\ref{fig:2d_densT1} and \ref{fig:2d_densT2}, in both the molecular and the quasiatomic states. The nature of the states does not alter qualitatively in Fig.~\ref{fig:2d_densT1}; they represent indeed the two molecular states, differing only by the bond length, etc. On the contrary, the nature of the molecular -- quasiatomic transition is very clear, since the density shown in Fig.~\ref{fig:2d_densT2}b splits with respect to (a) into two disjoint regions representing well separated states of atoms. The latter states are called quasiatomic because their size (cf. Fig.~\ref{fig:2d_zeta}) differs remarkably with respect to that of an isolated $H$ atom.

\begin{table*}[t!]
\caption{Values of the equilibrium lattice parameters, as well as the Mott-Hubbard and Mott criteria at both transitions.}
\label{tab:wf_coeffs2}
\resizebox{\textwidth}{!}{
 \begin{tabular}{c||r|r|r|r|r|r|}
 & $p_c (Ry/a_0^2)$ & $a_c (a_0)$ & $R_{eff,c} (a_0)$ & $\zeta_c^{-1} (a_0)$ & $(U/W)_c$ & $(\zeta_c a_c)^{-1}$\\\hline 
molecular I & $0.1102$ & 2.7626 & 1.1511 & 0.8571 & 1.3870 & 0.3103 \\\hline
molecular II& $0.1102$ & 2.6791 & 1.1881 & 1.0564 & 1.5778 & 0.3943 \\\hline\hline
molecular II& $0.1954$ & 2.4378 & 1.1296 & 1.0887 & 1.3112 & 0.4466 \\\hline
quasiatomic & $0.1954$ & 2.2313 & 1.9281 & 0.7398 & 0.6000 & 0.3316 \\ 
\end{tabular}
}
\end{table*}

The above evolution of electron density for molecules/atoms placed in the milieu of all other particles is supplemented with the selected relevant parameters displayed in Table~\ref{tab:wf_coeffs2}, where we list the values at the consecutive transitions (marked withe subscript $c$ in each case): $p=p_c$, $a=a_c$, $R=R_{eff,c}$, as well as provide the critical values of the Hubbard ration $(U/W)_c \sim 1$ and of the Mott criterion "$n_c^{1/D} a_B \sim 0.22$", here adopted to the two dimensional ($D=2$) case, for which the effective Bohr radius is $a_B \equiv \zeta_c^{-1}$ and the particle density $n_c = \tfrac{1}{a_c^2}$. Those three quantities are listed in the last two columns. It is amazing that those two sets of values, introduced via a rough estimates are not far off from the standard estimates \cite{Mott} at the transition to the quasiatomic state.

\begin{figure}
 \includegraphics[width=\linewidth]{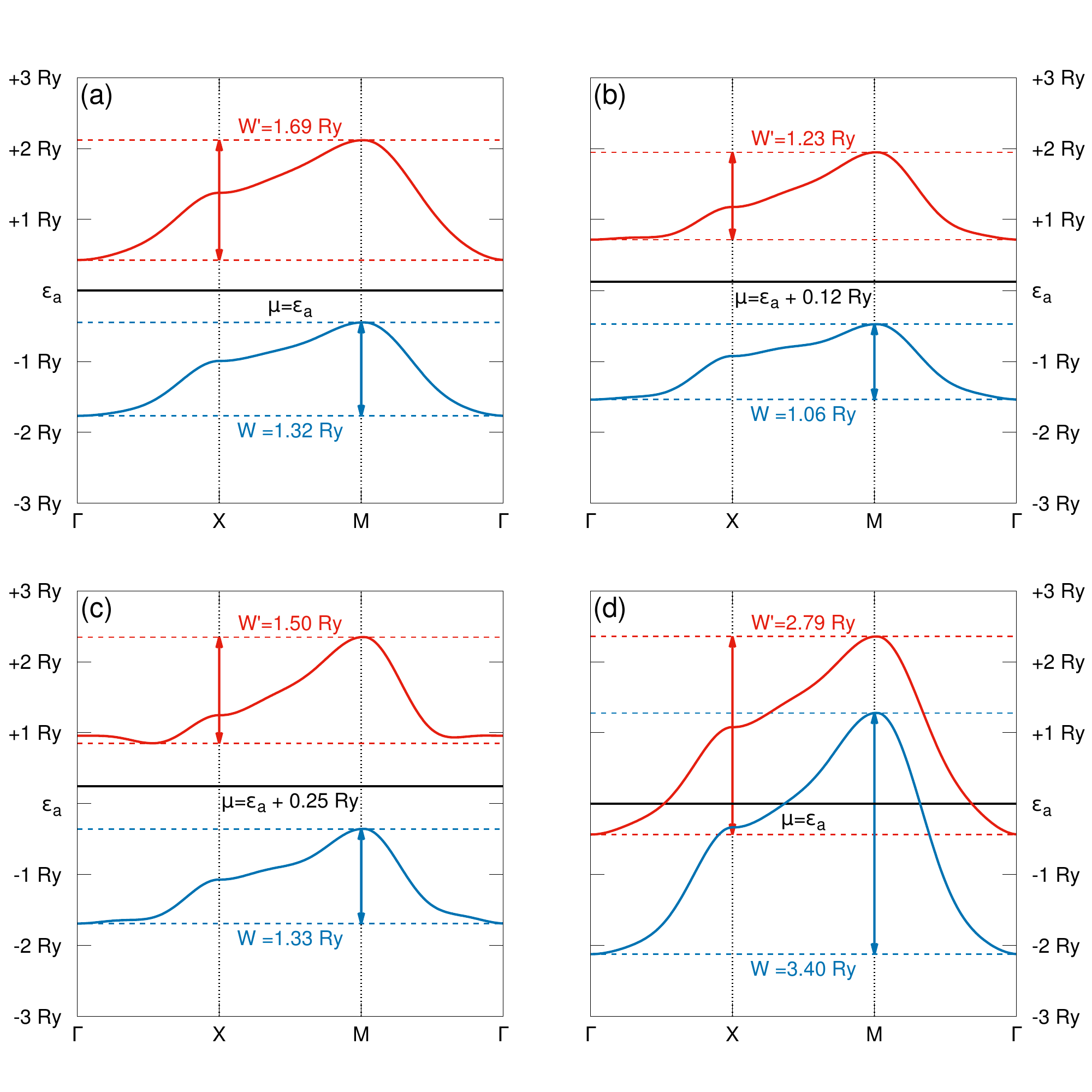}
 \caption{(a--b) Dispersion relations for bare bands at the molecular I \textrightarrow~molecular II transition ($p_{c1}=0.1102 Ry/a_0^2$). (c--d) the same at the molecular II \textrightarrow~quasiatomic transition ($p_{c2}=0.1954 Ry/a_0^2$). One expects that, the lowest band in Figs. (a--c) will split additionally into the  Hubbard subbands in those states as then $U/W > 1.5$, whereas in the state (d) they will overlap as the $U/W \approx 0.5$, i.e., the system eventually becomes a moderately correlated metal.
 }
 \label{fig:transition_dispersion}
\end{figure}

In Fig.~\ref{fig:transition_dispersion}a-b we have determined the two lowest bare-bands dispersion relations calculated with the renormalized hoppings parameters and at the transition from the molecular phase I to II, as well as in Fig.~\ref{fig:transition_dispersion}c-d at the transition from phase II to the quasiatomic phase. As the interactions are not included in those calculations, we do not have the Hubbard-subband structure at the transition from I to II. Nevertheless, since then $U/W > 1.5$, the structure represents that of an insulators, whereas the quasiatomic phase is metallic, as there is an appreciable band overlap is that state and the correlations are moderate to weak ($U/W \sim 0.5$). The phases I and II are both insulating; they differ only by different values of the microscopic parameters. It is tempting to suggest that while the phase I is diamagnetic, the phase II may be of insulating and  (antiferro)magnetic. However, this point requires a separate analysis.

\begin{figure}
 \includegraphics[width=\linewidth]{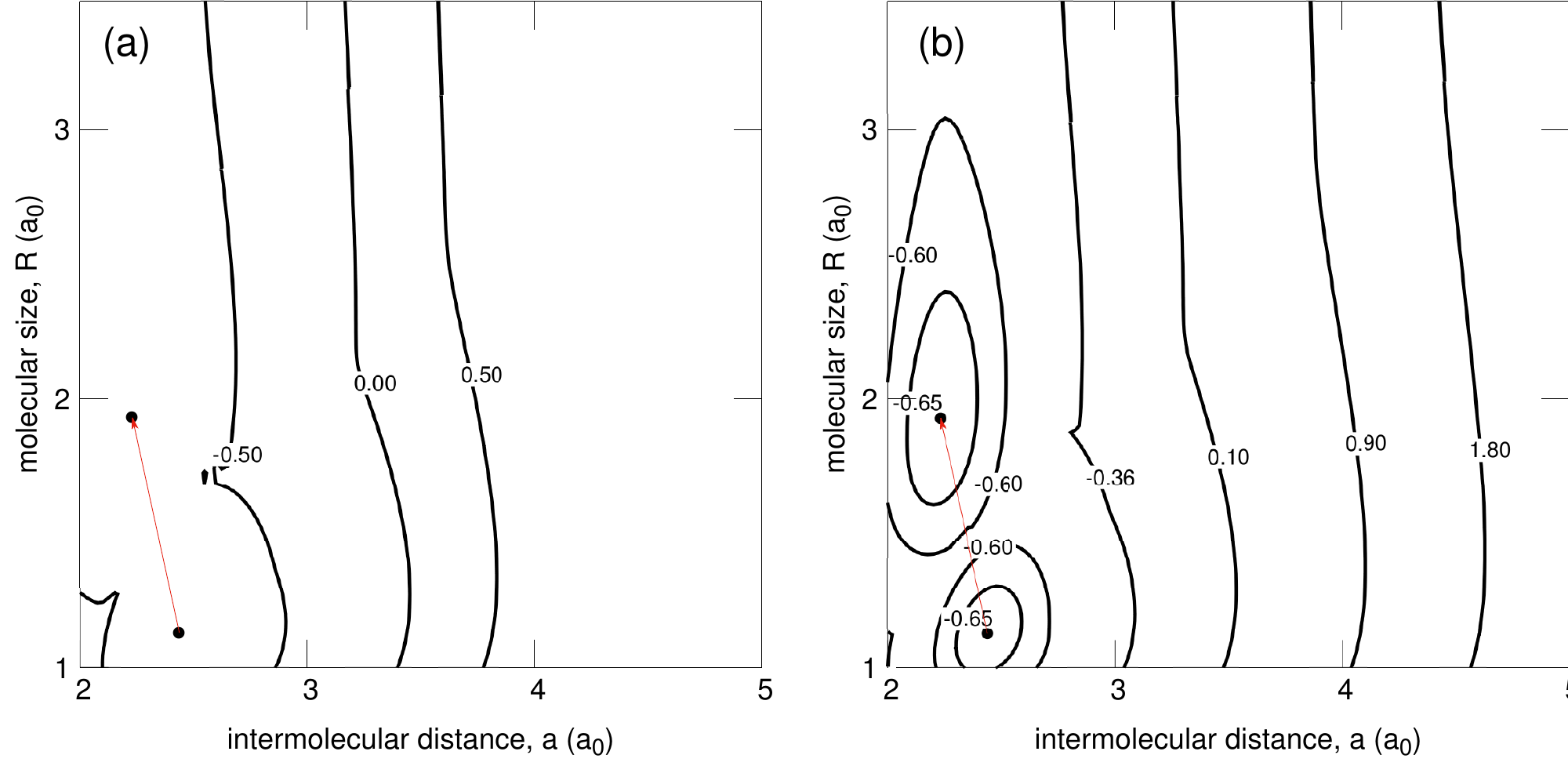}
 \caption{ Enthalpy isolines on the plane $a$-$R$ at the border between I and II states (a) and at the II-quasiatomic border (b). The points mark the minima with the arrow connecting them as a guide to the eye. Note that both transitions involve primarily a radical change in the effective molecule size $R_{eff}$.
 }
 \label{fig:transition_enthalpy}
\end{figure}

To provide an illustrative evidence for the existence of two distinct molecular phases, the corresponding to them enthalpy minima at those two transitions have been visualized in Fig.~\ref{fig:transition_enthalpy}a-b. We see that even in $2D$ there are two states and this circumstance may be regarded as a precursory effect for a number of such phases appearing in experiment on $3D$ systems \cite{Dalladay-Simpson,Drummond}.

\begin{figure*}
\includegraphics[width=.49\linewidth]{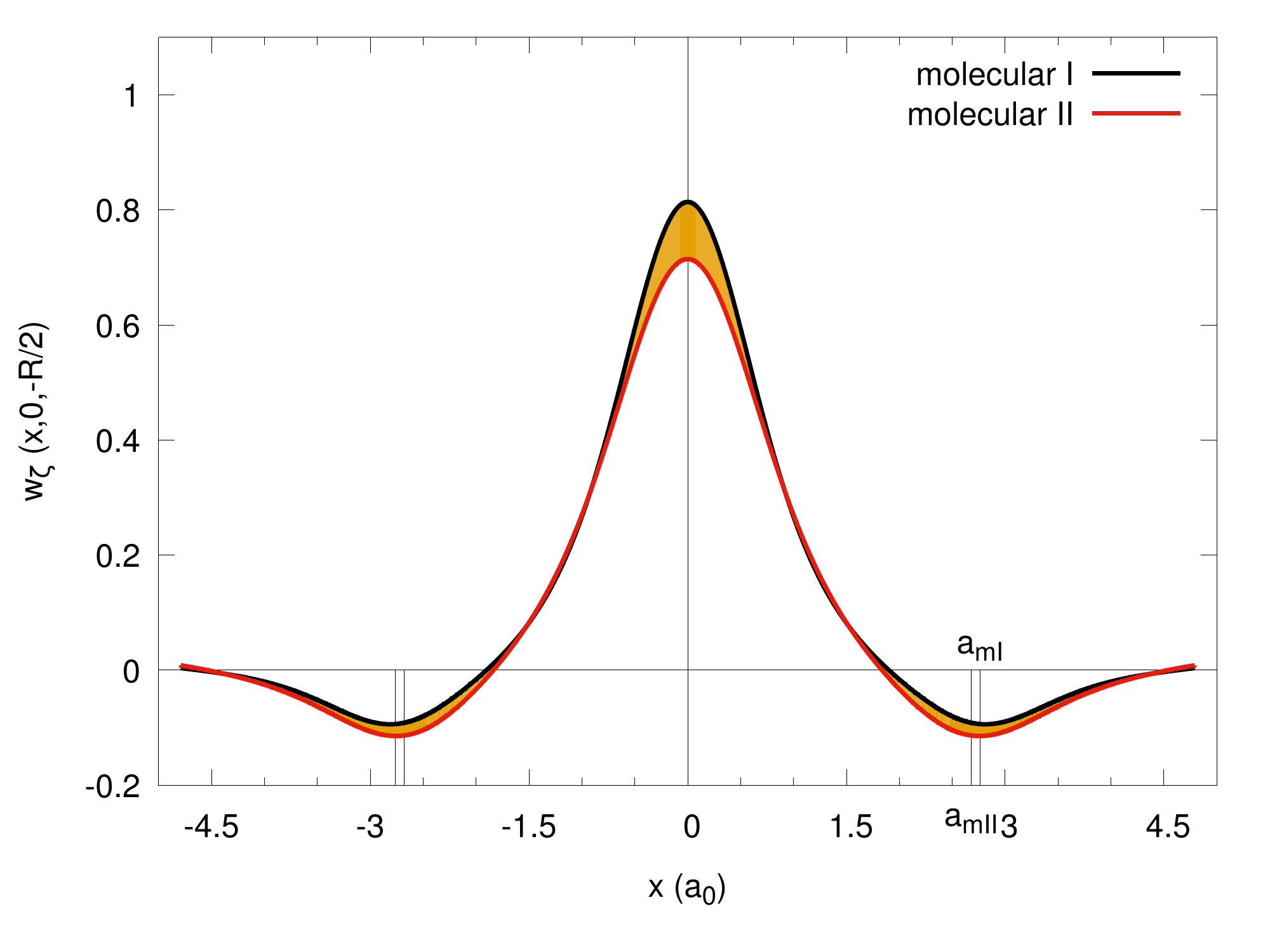}
\includegraphics[width=.49\linewidth]{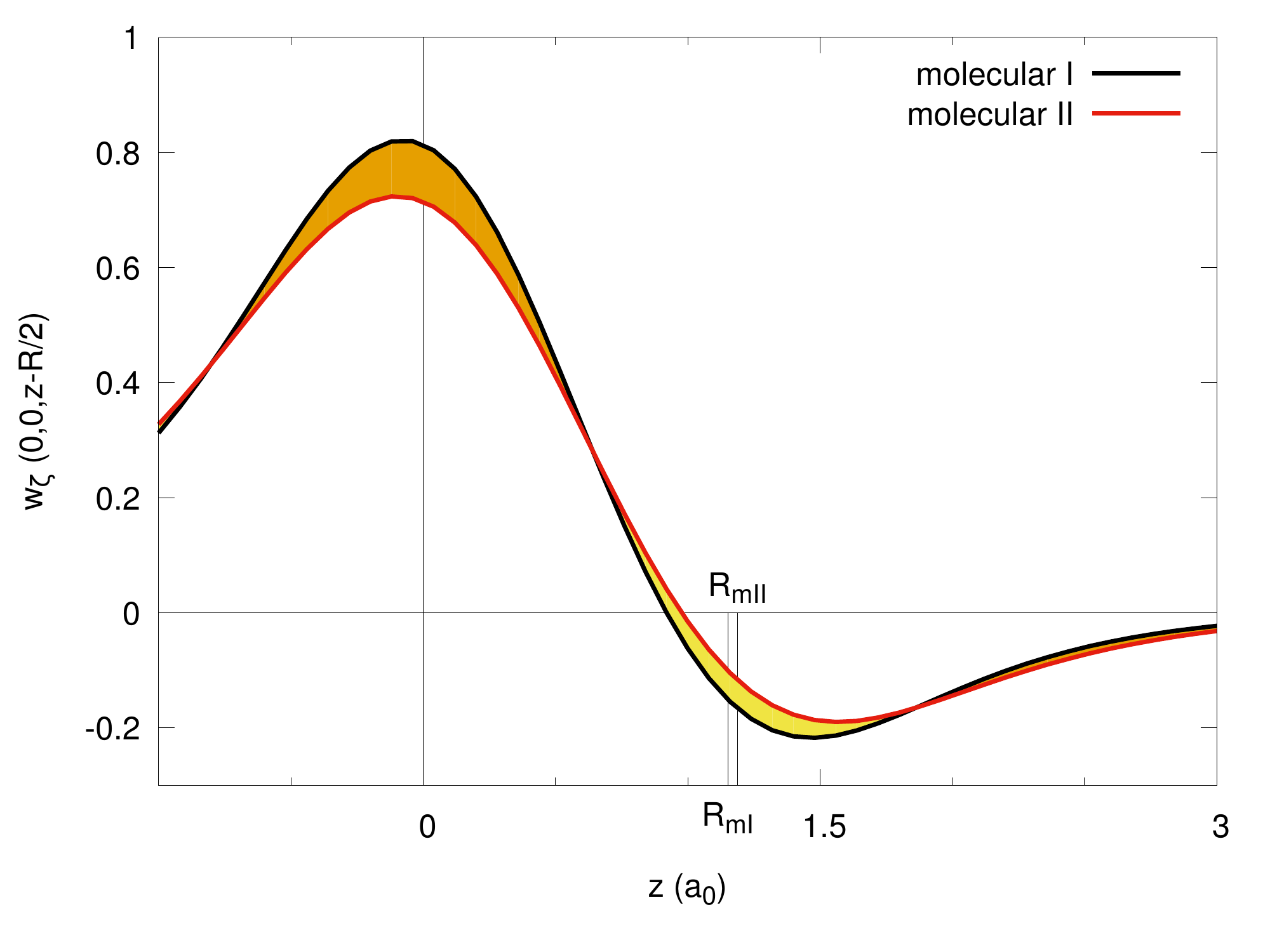}
\caption{Single-electron wave functions $w_{0}^{\beta}(\vec{r} = (x,0,0))$ (a) and $w_{0}^{\beta}(\vec{r} = (0,0,z-R/2))$ (b) (along the $z$ direction) in the  molecular phases I ($a=2.76261 (a_0)$, $R_{eff} = 1.1511 (a_0)$, $\zeta = 0.8571 (a_0^{-1})$)  II ($a=2.67911 (a_0)$, $R_{eff} = 1.1881 (a_0)$, $\zeta = 1.0564 (a_0^{-1})$) near the transition (at $p_{c1}=0.1102 Ry/a_0^2$).
}
\label{fig:wanniersT1}
\end{figure*}

\begin{figure*}
\includegraphics[width=.49\linewidth]{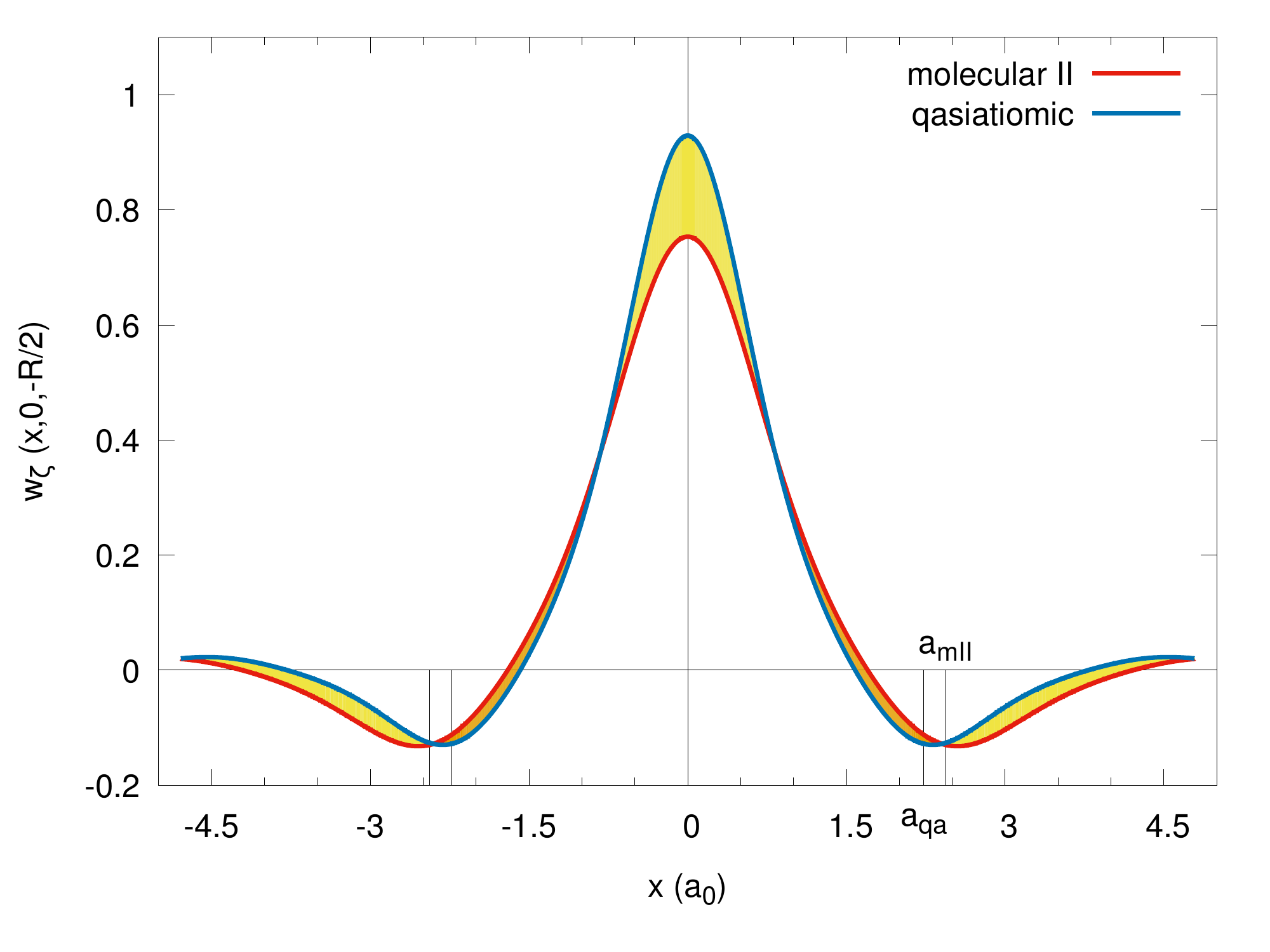}
\includegraphics[width=.49\linewidth]{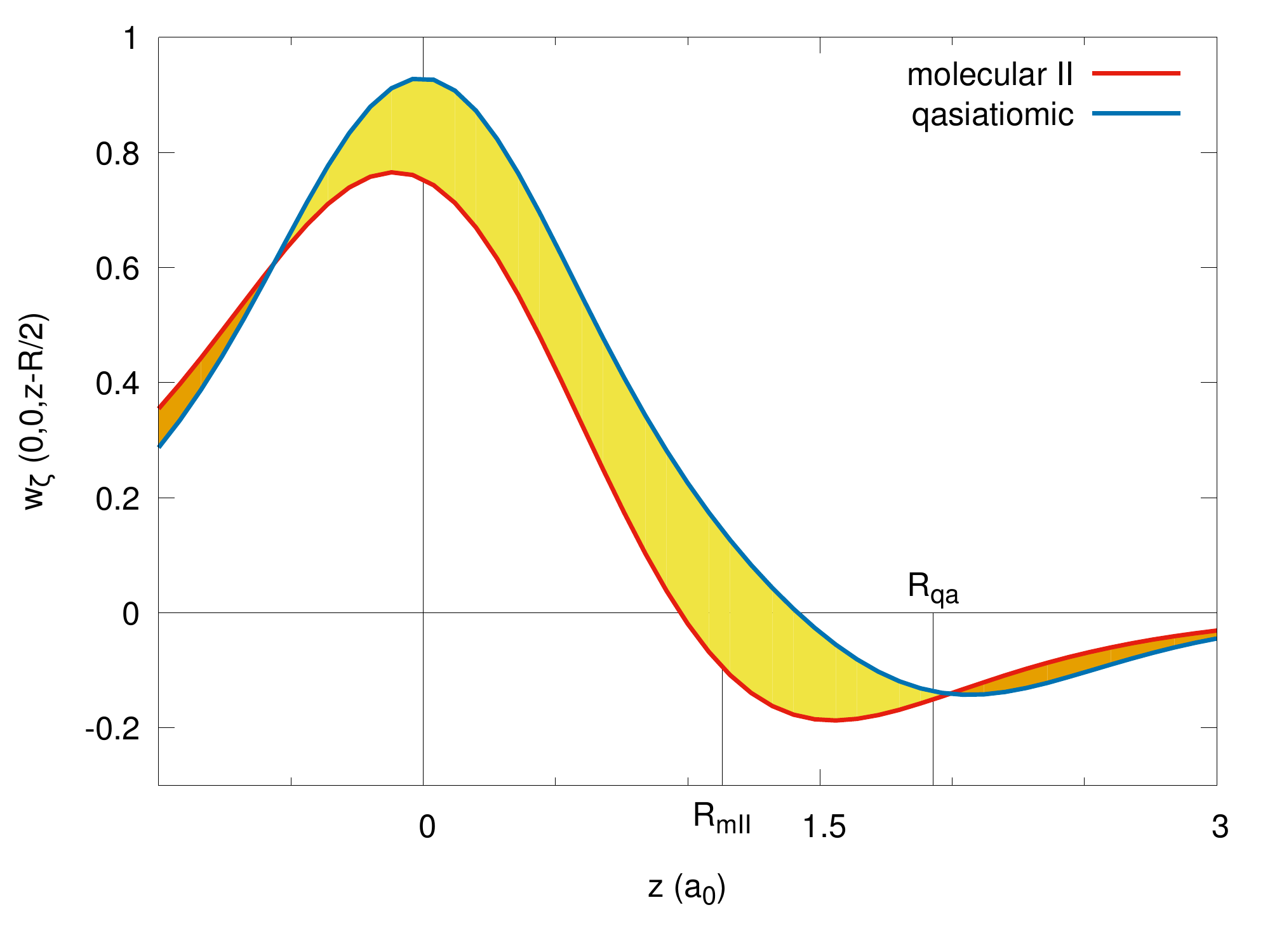}
\caption{Single-electron wave functions $w_{0}^{\beta}(\vec{r} = (x,0,0))$ (a) and $w_{0}^{\beta}(\vec{r} = (0,0,z-R/2))$ (b) (along the $z$ direction) in molecular phase II ($a=2.43783 (a_0)$, $R = 1.1296 (a_0)$, $\zeta = 1.0887 (a_0^{-1})$) and in the quasiatomic phase ($a=2.23133 (a_0)$, $R = 1.9281 (a_0)$, $\zeta = 0.7398 (a_0^{-1})$) near the transition (at $p_{c2}=0.1954 Ry/a_0^2$). Note the shrinking in the quasiatomic phase.
}
\label{fig:wanniersT2}
\end{figure*}

To illustrate the changes in the single-particle functions at the transitions, we have drawn the Wannier functions at the I\textrightarrow II- (cf. Fig.~\ref{fig:wanniersT1}) and II\textrightarrow quasiatomic-state (cf. Fig.~\ref{fig:wanniersT2}) transitions along in-plane ($x$ - (a)) and molecular ($z$ - (b)) directions. The two equilibrium lattice and bond parameters have been supplied in each of the Figures. Their evolution reflects perfectly the trend of the Slater-orbital size ($\zeta^{-1}$) jumps shown in Fig.~\ref{fig:2d_zeta}. It is amazing that they look more atomic-like in the last, metallic phase. However, the situation is not so simple, since at the same time the lattice parameter $a$ decreases appreciably in a discontinuous manner at the same time and therefore the change of Hamiltonian parameters is also influenced by that. Nonetheless, the bond-length changes are most important (cf. Fig.~\ref{fig:transition_enthalpy}.

Concluding this Section, the results presented in \cref{fig:2d_UW,fig:2d_W,fig:2d_avgs,fig:2d_densT1,fig:2d_densT2,fig:transition_dispersion,fig:transition_enthalpy,fig:wanniersT1,fig:wanniersT2} provide an unequivocally evidence for the molecular to quasiatomic phase transition at the critical pressure $p_{c2}=0.1954 Ry/a_0^2$. Obviously, a further evidence of metallicity in the latter phase would require a direct calculations of the electric conductivity. Namely, it would require an extension of the present approach to nonzero temperature, as here the conductivituy $\sigma_c$ at $T=0$ would take the values $\sigma_c=\infty$ in the molecular phases and $\sigma_c=\infty$ in the metallic ground state. However, the gap closure at the II \textrightarrow~quasiatomic discontinuous phase transition (cf. Fig.~\ref{fig:transition_dispersion}) provides a clear sign of metallicity in the latter phase.

\section{Outlook}
\label{sec:outlook}

\subsection{Brief summary and zero-point motion of atoms}
Let us summarize first our effort here. We have discussed the metallization of $2D$ stack of molecular hydrogen within the EDABI method. The method relies on an exact diagonalization of the Hamiltonian describing the dynamic processes within supercell containing $4H_2$ molecules; this cluster is subsequently repeated periodically in the both planar directions, with additional inclusion of the hopping and interelectronic interactions extending beyond the supercell (cf. Fig.~\ref{fig:hoppings_and_field}). In this respect, our approach represents a version of coupled-cluster approach \cite{Maska}. Furthermore, at each step of the iterative diagonalization procedure by Lanczos method we readjust the single-particle (Wannier) wave function until a fully microscopic ground-state energy configuration is reached. Therefore, the input  parameters are solely the atomic structure (cf. Fig.~\ref{fig:plane_model}) and  the finite  single-particle basis, limited here to $1s$ states. As a results we obtain the principal characteristics such as the lattice constant, the effective bond length, renormalized band structure and single-particle wavefunctions, and the ground-state enthalpy, all as a function of applied force. But first and foremost, we obtain the sequence of discontinuous phase transitions and in particular, the insulator -- metal transition from the $H_2$ insulator to $H$ metal. The atomization process is illustrated directly in Figs.~\ref{fig:2d_densT2}, where, the many-particle electron-density profiles have been drawn. All this provides the evidence that the hydrogen metallization represents a transition of the Mott-Hubbard type, though the starting material at ambient pressure is a diamagnetic (not antiferromagnetic) insulator. Hence, our approach represnts an essential extension of the concept of the canonical Mott-Hubbard transition.

Our analysis would be complete if we have supplemented the present work with the study of stability of the assumed protonic lattice in the metallic state. In other words, a separate question can be asked if the metallization is not associated with the transition to a liquid proton-electron plasma state \cite{Dzyabura,Weir,Tamblyn,Morales,Silvera}, though the corresponding transition in the liquid is also observed \cite{Nellis2015}. The latest experimental results \cite{Dias} support the view taken here that the lattice survives the strong first-order transitions (see however Ref.~\cite{Eremets}). The stability of the lattice can be justified by two features of our results. First, we have shown that at the transition the electron orbit shrinks remarkably (cf. \cref{fig:2d_zeta,fig:wanniersT2}), screening the charges on a short-distance scale and thus diminishing the repulsive energy. Second, in Appendix~\ref{app:B} we have estimated the amplitude and the energy contribution of the zero-point motion in the harmonic approximation \cite{Spalek,Kadzielawa3} and both in the inter- and intra-molecular driections. The inclusion of the zero-point motion can change only slightly the transition points without changing the overall features of our results. Note that the ZPM energy does not exceed $1\%$ of the total enthalpy value (cf. Fig.~\ref{fig:2d_enthalpy}), but up to about $4\%$ of the ground-state-energy value.

\subsection{Relation to other works}
The principal novelty of our approach relies on: (\emph{i}) implementing a combined first- and the second-quantization scheme which allow for a full \emph{ab-initio} analysis of the correlated state and without the appearance of the notorious double-counting problem; (\emph{ii}) determining the renormalized Wannier orbitals which supplement the whole picture qualitatively with respect to that obtained within the parametrized models; and (\emph{iii}) applying the concepts of the Mott-Hubbard transition to the canonical solid hydrogen system with $1s$ orbitals. The transition is accompanied by a simultaneous two-step transition from the correlated diamagnetic molecular insulator to the two-dimensional metal as a function of applied pressure. The inclusion of long-range Coulomb interaction should also be noted. The question remains as to whether such a bilayer system can be realized experimentally by e.g., covering a substrate with a plane of such stacked $H_2$ molecules, with the substrates of variable lattice parameter emulating the pressure applied to the system edges. Such an experiment could provide a direct realization of a bilayer crystal in a metallic state. In this case of a bilayer deposited on the substrate one would have to account also for the dynamics of the protons and electrons in the presence of a trapping them external (surface) potential which, if sufficiently strong, would suppress their zero-point vibrations.

Other theoretical works involve among others recent diffusion quantum Monte-Carlo simulations \cite{Azadi2,Drummond} and advanced DFT methods \cite{McMahon,Kudryashov}. Both methods predict the metallization for pressures in the range $400-500 GPa$ in the three-dimensional case. Here we show that the transition in the bilayer case is of the Mott-Hubbard type. The same type of the transition has been shown to exist in one-dimensional case \cite{Kadzielawa} for the molecular ladder. Therefore we expect that the same type of results can be expected in the $3D$ case, but a proof of that hypothesis requires more involved approach and should employ the incorporation of the Monte-Carlo methods into our scheme. The reason why our results are to certain degree independent of the lattice dimensionality is the fact that we include long-range Coulomb interactions between the electrons which make the results look more like those of mean field theory of those correlated fermion systems, results of which are only weakly dependent on the system dimensionality.  

\subsection{Concluding results}

So far we have analyzed the normal states only. As our bilayer metallic phase represents a moderate correlated system we can treat it as a bilayer system with correlation-driven pairing, analogously to our recent approach of the cuprates within the extended Hubbard model \cite{Spalek4,Zegrodnik,Kaczmarczyk}. However, the situation is not that simple as here the electron-lattice interactions can be quite strong, as one can see already on the $H_2$ and $(H_2)_2$ examples \cite{Kadzielawa3}. In effect, both the correlation and electron-lattice parts should be treated on equal footing. In that case, one can estimate their relative contributions to the superconducting critical temperature and in such a manner complement the estimates based purely on the electron-lattice contribution \cite{Kudryashov,Ashcroft,Borinaga}. We should be able to see a progress along these lines in the near future.

\section{Acknowledgments}
\label{sec:ackn}

The work was financially supported by the National Science Centre (NCN), through Grant {No.~DEC-2012/04/A/ST3/00342}. The computations have been performed in part on the supercomputer TERA-ACMiN AGH and partly on the supercomputer EDABI located at the Jagiellonian University.

\appendix
\section{Bare bandwidth \emph{W} of the electrons in the correlated state}
\label{app:A}

To compute $U/W$ ratio, the \emph{bandwidth} $W$ can be obtained from diagonalization of the single--electron part of Hamiltonian \eqref{eq:hamiltonian_tot}, i.e.,
\begin{align}
 \label{eq:hamiltonian_sinel}
 &\hat{\mathcal{H}} = \sum\limits_{i\mu\sigma}\epsilon_i^{\mu}\NUM{i}{2}+\sideset{}{'}\sum\limits_{ij\mu\nu\sigma}t_{ij}^{\mu\nu}\hat{c}_{i\mu,\sigma}^{\dagger}\hat{c}_{j\nu,\sigma}^{}
\end{align}
 In accordance with the translational invariance of the system in the \emph{x-y} plane, (~\ref{eq:hamiltonian_sinel}) can be rewritten in the momentum ($\vec{k}$) representation in the form
\begin{align}
 \label{eq:hamiltonian_sinelmom}
 \hat{\mathcal{H}} =& \sum\limits_{\vec{k}\mu\sigma}\epsilon_{}^{\mu}\NUM{\vec{k}}{2}+\sideset{}{'}\sum\limits_{\vec{k}\mu\nu\sigma}\hat{c}_{\vec{k}\mu,\sigma}^{\dagger}\hat{c}_{\vec{k}\nu,\sigma}^{}\sum\limits_{l}^{ }t_{p(l,\mu)q(l,\nu)}^{\mu\nu}\text{exp}\big(-i\vec{k}\vec{r}_l^{\mu\nu}\big) \\\notag
                   =& \sum\limits_{\vec{k}\mu}\epsilon_{}^{\mu}\NUM{\vec{k}}{0}+\sideset{}{'}\sum\limits_{\vec{k}\mu\nu\sigma}\hat{c}_{\vec{k}\mu,\sigma}^{\dagger}\hat{c}_{\vec{k}\nu,\sigma}^{}\sum\limits_{l}^{ }Z^{\mu\nu}_{\vec{k}},
\end{align}
where primed summation refers to $\mu \neq \nu$, index $l$ enumerates molecules in the assumed neighborhood, i.e., $\vec{r}_l^{\mu\nu} = \vec{R}_{0}^{\mu} - \vec{R}_{l}^{\nu}$. The functions $p(l,\mu)$ and $q(l,\nu)$ map $l,\mu,\nu$ to proper indexing of the hoppings.  
Note that one may select $\vec{k} = (\tfrac{2\pi n}{a_x},\tfrac{2\pi m}{a_y}) = (\tfrac{2\pi n}{a},\tfrac{2\pi m}{a})$ where $m,n$ are integers, as we did in our considerations. In effect, the single-electron Hamiltonian can be recast in the matrix form for each spin, i.e.,
\begin{align}
 \label{eq:hamiltonian_sinmatrix}
 &\hat{\mathcal{H}}_{\sigma} = \begin{pmatrix}\hat{c}_{\vec{k}\alpha,\sigma}^{\dagger}& \hat{c}_{\vec{k}\beta,\sigma}^{\dagger}\end{pmatrix} \begin{pmatrix} \epsilon_{\vec{k}}^{\alpha}&Z^{\alpha\beta}_{\vec{k}}\\Z^{\beta\alpha}_{\vec{k}}&\epsilon_{\vec{k}}^{\beta} \end{pmatrix}\begin{pmatrix}\hat{c}_{\vec{k}\alpha,\sigma}^{}\\\hat{c}_{\vec{k}\beta,\sigma}^{}\end{pmatrix}\equiv\mathbf{c}^{\dagger}\,\mathbb{H}\,\mathbf{c}.
\end{align}
Diagonalization of matrix $\mathbb{H}$ provides the bare dispersion relation $\tilde{\epsilon}(\vec{k})$. For our $2D$ molecular crystal two, spin-degenerate, branches $\tilde{\epsilon}^{up}(\vec{k})$ and $\tilde{\epsilon}^{down}(\vec{k})$ appear.
The matrix $\mathbb{H}$ is constructed in a straightforward manner for particular $\vec{k}$, i.e., by computing numerically $Z_{k}^{\mu\nu}$ with $\vec{r}_l^{\mu\nu}$ up to the $13th$ coordination zone. Subsequently, $\mathbb{H}$ is diagonalized and the two eigenvalues $\tilde{\epsilon}^{up}(\vec{k})$, $\tilde{\epsilon}^{down}(\vec{k})$ are obtained.
For the \emph{half filling} considered here, only $\tilde{\epsilon}^{down}$ band is occupied by electrons. Therefore $W$ is defined in a standard manner, i.e.,
\begin{align}
  \label{eq:def_w}
  &W = \tilde{\epsilon}^{down}_{max} - \tilde{\epsilon}^{down}_{min}.
\end{align}
Both the maximal and the minimal values, $\tilde{\epsilon}^{down}_{max}$ and $\tilde{\epsilon}^{down}_{min}$, are obtained numerically for $\vec{k}$ by scanning the eigenvalues in the first Brillouin zone. Those values were used when plotting $U-W$ and $U/W$ in Figs.~\ref{fig:2d_UW} and ~\ref{fig:2d_W} in main text. Note that in the molecular state the lower band is nominally filled, whereas in the metallic state the bands overlap.

\section{Assessment of zero-point motion in harmonic approximation}
\label{app:B}

We estimate the zero-point motion (ZPM) of our system by introducing a ion position uncertainty
\begin{align}
  \delta \vec{r} \equiv (\delta x,\delta y,\delta z),
\end{align}
then by splitting the problem into two parts: \emph{(i) ZPM in a molecule} ($ \delta \vec{r}_{i} \equiv (0,0,\delta z)$), \emph{(ii) ZPM of a molecule in the crystal field} ($ \delta \vec{r}_{ii} \equiv (\delta x,\delta y,0)$). In both cases the kinetic energy of the $H_2$ molecule is
\begin{align}
  E_{kin} \equiv 2 \frac{\hbar^2 \delta \vec{p}^2}{2 M},
\end{align}
where $\hbar \stackrel{a.u.}{\equiv} 1$, $M \stackrel{a.u.}{\equiv} 1836.15267247 \times\tfrac{1}{2}$, and $\delta \vec{p}$ is approximated via the Heisenberg uncertainty principle
\begin{align}
\delta \vec{p}^2\delta \vec{r}^2 \leq \frac{3 \hbar^2}{4} \ \ \ \stackrel{est.}{\rightarrow} \ \ \ \delta \vec{p}^2 = \frac{3 \hbar^2}{4 \delta \vec{r}^2},
\end{align}
hence $E_{kin} = \tfrac{3 \hbar^4}{4M \delta \vec{r}^2}$.

The potential energy is calculated separately for the cases \emph{(i)} and \emph{(ii)}.

\subsection{ZPM for \texorpdfstring{$H_2$}{H2} molecule}

We base our approach by our earlier work \cite{Spalek,Kadzielawa3}. We define the potential
\begin{align}
V_{m} (R,\delta \vec{r}_{i}) \equiv E(R + \delta z) - E_B,
\end{align}
where $E(R)$ is the energy of the molecule of the molecular size $R$, and $E_B \equiv E(1.43042 a_0)$, the minimum of energy for static molecule (so that our potential used static equilibrium as a reference point).

The energy gain from the ionic movement is given by an expression
\begin{align}
  \label{eq:enGainMol}
  \Delta E(R,\delta \vec{r}_{i}) &= E_{kin}(\delta \vec{r}_{i}) + V_{m}(R,\delta \vec{r}_{i}) \\\notag
  &= \frac{3 \hbar^4}{4M \delta z^2} + E(R + \delta z) - E_B.
\end{align}
For the given molecular size $R$ we minimize expression \eqref{eq:enGainMol} with respect to $\delta z$.

\subsection{ZPM per molecule in the crystal}
For the case of the molecule in the crystal field we assume that the electrons do not contribute to the ionic potential, hence
\begin{align}
&V_{crystal} (a,R,\delta \vec{r}_{i}) \equiv \\\notag
&\sum_{\text{interaction cell}} \frac{e^2}{| \vec{R}_i(a) + (0,0,-R/2) - \delta\vec{r} |} \\\notag
+&\sum_{\text{interaction cell}} \frac{e^2}{| \vec{R}_i(a) + (0,0,R/2) - \delta\vec{r} |} \\\notag
-& V_{static}(a,R),
\end{align}
where $a$ is the intermolecular distance, $R$ is the molecule size, $e \stackrel{a.u.}{\equiv} \sqrt{2}$ is the charge of a hydrogen ion, \emph{interaction cell} refers to the molecules we considered as our background (cf. Fig.~\ref{fig:hoppings_and_field} for the background considered in this paper) at the positions $\vec{R}_i(a)$, and $V_{static}(a,R)$ is the potential of static molecules
\begin{align}
V_{static} (a,R) \equiv& \sum_{\text{interaction cell}} \frac{e^2}{| \vec{R}_i(a) + (0,0,-R/2) |} \\\notag
+& \sum_{\text{interaction cell}} \frac{e^2}{| \vec{R}_i(a) + (0,0,R/2) |}.
\end{align}

The energy gain from the ionic movement is given by an expression
\begin{align}
  \label{eq:enGainCry}
  \Delta E(R,\delta \vec{r}_{ii}) =& E_{kin}(\delta \vec{r}_{ii}) + V_{crystal}(a,R,\delta \vec{r}_{ii}) = \\\notag
  &\frac{3 \hbar^4}{4M (\delta x^2 + \delta y^2)} + V_{crystal}(a,R,(\delta x, \delta y,0)).
\end{align}
For the given intermolecular distance $a$ and molecular size $R$ we minimize expression \eqref{eq:enGainCry} with respect to $\delta x$ and $\delta y$.

\subsection{Numerical results at the transitions}

In Table~\ref{tab:ZPM} we present both absolute and relative magnitude of ZPM, as well as all the possible modes with their corresponding energies (in Rydberg per molecule).

\begin{table*}[t!]
\caption{Magnitude of the zero-point motion and all possible modes at the transitions and for the ambient pressure ($p=0$). Energy values are in Rydbergs per molecule.}
\label{tab:ZPM}
\resizebox{\textwidth}{!}{
 \begin{tabular}{c||r|r|r|r||r|r||}
 $p (Ry/a_0^2)$ & phase & $a (a_0)$ & $R_{eff} (a_0)$ & $E_G (Ry)$ & $E_{mode} (Ry)$ & direction of the mode \\\hline 
   $0$& molecular I & 4.3371 & 1.4031 & -2.3858 & $ 2 \times 0.01605$ & $(\pm \tfrac{1}{\sqrt{2}}, \pm \tfrac{1}{\sqrt{2}},0)$ or $(\pm \tfrac{1}{\sqrt{2}}, \mp \tfrac{1}{\sqrt{2}},0)$\\\cline{2-7}
     &     \mc{2}{r}{ } &  \mc{2}{r||}{ } & $ 2 \times 0.01608$ & $(\pm 1,0,0)$ or $(0,\pm 1,0)$\\\cline{6-7} 
     &     \mc{2}{r}{ } &  \mc{2}{r||}{ } & 0.01837 & $(0,0,1)$\\\cline{6-7} 
     &     \mc{2}{r}{ } &  \mc{2}{r}{$\mathbf{E_{ZPM}=}$} & \mc{1}{r}{ \bf0.08263 Ry} & $\mathbf{|E_{ZPM}|/|E_G| = 3.46\%}$\\\hline\hline
$0.1102$ & molecular I & 2.7626 & 1.1511 & -2.0674 & $ 2 \times 0.03035$ & $(\pm \tfrac{1}{\sqrt{2}}, \pm \tfrac{1}{\sqrt{2}},0)$ and $(\pm \tfrac{1}{\sqrt{2}}, \mp \tfrac{1}{\sqrt{2}},0)$\\\cline{2-7} 
     &     \mc{2}{r}{ } &  \mc{2}{r||}{ } & $ 2 \times 0.03044$ & $(\pm 1,0,0)$ and $(0,\pm 1,0)$\\\cline{6-7} 
     &     \mc{2}{r}{ } &  \mc{2}{r||}{ } & 0.00452 & $(0,0,1)$\\\cline{6-7} 
     &     \mc{2}{r}{ } &  \mc{2}{r}{$\mathbf{E_{ZPM}=}$} & \mc{1}{r}{ \bf0.1261 Ry} & $\mathbf{|E_{ZPM}|/|E_G| = 6.10\%}$\\\cline{2-7}
     & molecular II & 2.6791 & 1.1881 & -2.0173 & $ 2 \times 0.03140$ & $(\pm \tfrac{1}{\sqrt{2}}, \pm \tfrac{1}{\sqrt{2}},0)$ and $(\pm \tfrac{1}{\sqrt{2}}, \mp \tfrac{1}{\sqrt{2}},0)$\\\cline{2-7} 
     &     \mc{2}{r}{ } &  \mc{2}{r||}{ } & $ 2 \times 0.03150$ & $(\pm 1,0,0)$ and $(0,\pm 1,0)$\\\cline{6-7} 
     &     \mc{2}{r}{ } &  \mc{2}{r||}{ } & 0.00557 & $(0,0,1)$\\\cline{6-7} 
     &     \mc{2}{r}{ } &  \mc{2}{r}{$\mathbf{E_{ZPM}=}$} & \mc{1}{r}{ \bf0.13137 Ry} & $\mathbf{|E_{ZPM}|/|E_G| = 6.51\%}$\\\hline\hline
$0.1954$ & molecular II & 2.4378 & 1.1296 & -1.8362 & $ 2 \times 0.03584$ & $(\pm \tfrac{1}{\sqrt{2}}, \pm \tfrac{1}{\sqrt{2}},0)$ and $(\pm \tfrac{1}{\sqrt{2}}, \mp \tfrac{1}{\sqrt{2}},0)$\\\cline{2-7} 
     &     \mc{2}{r}{ } &  \mc{2}{r||}{ } & $ 2 \times 0.03596$ & $(\pm 1,0,0)$ and $(0,\pm 1,0)$\\\cline{6-7} 
     &     \mc{2}{r}{ } &  \mc{2}{r||}{ } & 0.00402 & $(0,0,1)$\\\cline{6-7} 
     &     \mc{2}{r}{ } &  \mc{2}{r}{$\mathbf{E_{ZPM}=}$} & \mc{1}{r}{ \bf0.14762 Ry} & $\mathbf{|E_{ZPM}|/|E_G| = 8.04\%}$\\\cline{2-7}
     & quasiatomic & 2.2313 & 1.9281 & -1.6478 & $ 2 \times 0.03478$ & $(\pm \tfrac{1}{\sqrt{2}}, \pm \tfrac{1}{\sqrt{2}},0)$ and $(\pm \tfrac{1}{\sqrt{2}}, \mp \tfrac{1}{\sqrt{2}},0)$\\\cline{2-7} 
     &     \mc{2}{r}{ } &  \mc{2}{r||}{ } & $ 2 \times 0.03493$ & $(\pm 1,0,0)$ and $(0,\pm 1,0)$\\\cline{6-7} 
     &     \mc{2}{r}{ } &  \mc{2}{r||}{ } & 0.00162 & $(0,0,1)$\\\cline{6-7} 
     &     \mc{2}{r}{ } &  \mc{2}{r}{$\mathbf{E_{ZPM}=}$} & \mc{1}{r}{ \bf0.14104 Ry} & $\mathbf{|E_{ZPM}|/|E_G| = 8.56\%}$\\\hline\hline
\end{tabular}
}
\end{table*}
\clearpage
\bibliography{}

\begin{thebibliography}{45}%
\makeatletter
\providecommand \@ifxundefined [1]{%
 \@ifx{#1\undefined}
}%
\providecommand \@ifnum [1]{%
 \ifnum #1\expandafter \@firstoftwo
 \else \expandafter \@secondoftwo
 \fi
}%
\providecommand \@ifx [1]{%
 \ifx #1\expandafter \@firstoftwo
 \else \expandafter \@secondoftwo
 \fi
}%
\providecommand \natexlab [1]{#1}%
\providecommand \enquote  [1]{``#1''}%
\providecommand \bibnamefont  [1]{#1}%
\providecommand \bibfnamefont [1]{#1}%
\providecommand \citenamefont [1]{#1}%
\providecommand \href@noop [0]{\@secondoftwo}%
\providecommand \href [0]{\begingroup \@sanitize@url \@href}%
\providecommand \@href[1]{\@@startlink{#1}\@@href}%
\providecommand \@@href[1]{\endgroup#1\@@endlink}%
\providecommand \@sanitize@url [0]{\catcode `\\12\catcode `\$12\catcode
  `\&12\catcode `\#12\catcode `\^12\catcode `\_12\catcode `\%12\relax}%
\providecommand \@@startlink[1]{}%
\providecommand \@@endlink[0]{}%
\providecommand \url  [0]{\begingroup\@sanitize@url \@url }%
\providecommand \@url [1]{\endgroup\@href {#1}{\urlprefix }}%
\providecommand \urlprefix  [0]{URL }%
\providecommand \Eprint [0]{\href }%
\providecommand \doibase [0]{http://dx.doi.org/}%
\providecommand \selectlanguage [0]{\@gobble}%
\providecommand \bibinfo  [0]{\@secondoftwo}%
\providecommand \bibfield  [0]{\@secondoftwo}%
\providecommand \translation [1]{[#1]}%
\providecommand \BibitemOpen [0]{}%
\providecommand \bibitemStop [0]{}%
\providecommand \bibitemNoStop [0]{.\EOS\space}%
\providecommand \EOS [0]{\spacefactor3000\relax}%
\providecommand \BibitemShut  [1]{\csname bibitem#1\endcsname}%
\let\auto@bib@innerbib\@empty
\bibitem [{\citenamefont {Ko\l{}os}\ and\ \citenamefont
  {Wolniewicz}(1968)}]{Kolos}%
  \BibitemOpen
  \bibfield  {author} {\bibinfo {author} {\bibfnamefont {W.}~\bibnamefont
  {Ko\l{}os}}\ and\ \bibinfo {author} {\bibfnamefont {L.}~\bibnamefont
  {Wolniewicz}},\ }\bibfield  {title} {\enquote {\bibinfo {title} {{Improved
  Theoretical Ground-State Energy of the Hydrogen Molecule}},}\ }\href
  {\doibase 10.1063/1.1669836} {\bibfield  {journal} {\bibinfo  {journal} {J.
  Chem. Phys.}\ }\textbf {\bibinfo {volume} {49}},\ \bibinfo {pages} {404}
  (\bibinfo {year} {1968})}\BibitemShut {NoStop}%
\bibitem [{\citenamefont {Pachucki}\ and\ \citenamefont
  {Komasa}(2016)}]{Pachucki}%
  \BibitemOpen
  \bibfield  {author} {\bibinfo {author} {\bibfnamefont {K.}~\bibnamefont
  {Pachucki}}\ and\ \bibinfo {author} {\bibfnamefont {J.}~\bibnamefont
  {Komasa}},\ }\bibfield  {title} {\enquote {\bibinfo {title} {{Schr\"odinger
  equation solved for the hydrogen molecule with unprecedented accuracy}},}\
  }\href {\doibase 10.1063/1.4948309} {\bibfield  {journal} {\bibinfo
  {journal} {J. Chem. Phys.}\ }\textbf {\bibinfo {volume} {144}},\ \bibinfo
  {pages} {164306} (\bibinfo {year} {2016})}\BibitemShut {NoStop}%
\bibitem [{\citenamefont {Dalladay-Simpson}\ \emph {et~al.}(2016)\citenamefont
  {Dalladay-Simpson}, \citenamefont {Howie},\ and\ \citenamefont
  {Gregoryanz}}]{Dalladay-Simpson}%
  \BibitemOpen
  \bibfield  {author} {\bibinfo {author} {\bibfnamefont {Ph.}\ \bibnamefont
  {Dalladay-Simpson}}, \bibinfo {author} {\bibfnamefont {R.~T.}\ \bibnamefont
  {Howie}}, \ and\ \bibinfo {author} {\bibfnamefont {E.}~\bibnamefont
  {Gregoryanz}},\ }\bibfield  {title} {\enquote {\bibinfo {title} {{Evidence
  for a new phase of dense hydrogen above 325 gigapascals}},}\ }\href {\doibase
  10.1038/nature16164} {\bibfield  {journal} {\bibinfo  {journal} {Nature}\
  }\textbf {\bibinfo {volume} {529}},\ \bibinfo {pages} {63} (\bibinfo {year}
  {2016})}\BibitemShut {NoStop}%
\bibitem [{\citenamefont {Dzyabura}\ \emph {et~al.}(2013)\citenamefont
  {Dzyabura}, \citenamefont {Zaghoo},\ and\ \citenamefont
  {Silvera}}]{Dzyabura}%
  \BibitemOpen
  \bibfield  {author} {\bibinfo {author} {\bibfnamefont {V.}~\bibnamefont
  {Dzyabura}}, \bibinfo {author} {\bibfnamefont {M.}~\bibnamefont {Zaghoo}}, \
  and\ \bibinfo {author} {\bibfnamefont {I.~F.}\ \bibnamefont {Silvera}},\
  }\bibfield  {title} {\enquote {\bibinfo {title} {{Evidence of a
  liquid–liquid phase transition in hot dense hydrogen}},}\ }\href {\doibase
  10.1073/pnas.1300718110} {\bibfield  {journal} {\bibinfo  {journal} {Proc.
  Natl. Acad. Sci.}\ }\textbf {\bibinfo {volume} {110}},\ \bibinfo {pages}
  {8040} (\bibinfo {year} {2013})}\BibitemShut {NoStop}%
\bibitem [{\citenamefont {Howie}\ \emph {et~al.}(2015)\citenamefont {Howie},
  \citenamefont {Dalladay-Simpson},\ and\ \citenamefont {Gregoryanz}}]{Howie}%
  \BibitemOpen
  \bibfield  {author} {\bibinfo {author} {\bibfnamefont {R.~T.}\ \bibnamefont
  {Howie}}, \bibinfo {author} {\bibfnamefont {Ph.}\ \bibnamefont
  {Dalladay-Simpson}}, \ and\ \bibinfo {author} {\bibfnamefont
  {E.}~\bibnamefont {Gregoryanz}},\ }\bibfield  {title} {\enquote {\bibinfo
  {title} {{Evidence for a new phase of dense hydrogen above 325
  gigapascals}},}\ }\href {\doibase 10.1038/nmat4213} {\bibfield  {journal}
  {\bibinfo  {journal} {Nat. Mat}\ }\textbf {\bibinfo {volume} {14}},\ \bibinfo
  {pages} {495} (\bibinfo {year} {2015})}\BibitemShut {NoStop}%
\bibitem [{\citenamefont {Dias}\ and\ \citenamefont {Silvera}(2017)}]{Dias}%
  \BibitemOpen
  \bibfield  {author} {\bibinfo {author} {\bibfnamefont {R.~P.}\ \bibnamefont
  {Dias}}\ and\ \bibinfo {author} {\bibfnamefont {I.~F.}\ \bibnamefont
  {Silvera}},\ }\bibfield  {title} {\enquote {\bibinfo {title} {{Observation of
  the Wigner-Huntington transition to metallic hydrogen}},}\ }\href {\doibase
  10.1126/science.aal1579} {\bibfield  {journal} {\bibinfo  {journal}
  {Science}\ } (\bibinfo {year} {2017}),\ 10.1126/science.aal1579}\BibitemShut
  {NoStop}%
\bibitem [{\citenamefont {McMinis}\ \emph {et~al.}(2015)\citenamefont
  {McMinis}, \citenamefont {Clay}, \citenamefont {Lee},\ and\ \citenamefont
  {Morales}}]{McMinis}%
  \BibitemOpen
  \bibfield  {author} {\bibinfo {author} {\bibfnamefont {J.}~\bibnamefont
  {McMinis}}, \bibinfo {author} {\bibfnamefont {R.~C.}\ \bibnamefont {Clay}},
  \bibinfo {author} {\bibfnamefont {D.}~\bibnamefont {Lee}}, \ and\ \bibinfo
  {author} {\bibfnamefont {M.~A.}\ \bibnamefont {Morales}},\ }\bibfield
  {title} {\enquote {\bibinfo {title} {{Molecular to Atomic Phase Transition in
  Hydrogen under High Pressure}},}\ }\href {\doibase
  10.1103/PhysRevLett.114.105305} {\bibfield  {journal} {\bibinfo  {journal}
  {Phys. Rev. Lett.}\ }\textbf {\bibinfo {volume} {114}},\ \bibinfo {pages}
  {105305} (\bibinfo {year} {2015})}\BibitemShut {NoStop}%
\bibitem [{\citenamefont {Drummond}\ \emph {et~al.}(2015)\citenamefont
  {Drummond}, \citenamefont {Monserrat}, \citenamefont {Lloyd-Williams},
  \citenamefont {L\'{o}pez~R\'{i}os}, \citenamefont {Pickard},\ and\
  \citenamefont {Needs}}]{Drummond}%
  \BibitemOpen
  \bibfield  {author} {\bibinfo {author} {\bibfnamefont {N.~D.}\ \bibnamefont
  {Drummond}}, \bibinfo {author} {\bibfnamefont {B.}~\bibnamefont {Monserrat}},
  \bibinfo {author} {\bibfnamefont {J.~H.}\ \bibnamefont {Lloyd-Williams}},
  \bibinfo {author} {\bibfnamefont {P.}~\bibnamefont {L\'{o}pez~R\'{i}os}},
  \bibinfo {author} {\bibfnamefont {Ch.~J.}\ \bibnamefont {Pickard}}, \ and\
  \bibinfo {author} {\bibfnamefont {R.~J.}\ \bibnamefont {Needs}},\ }\bibfield
  {title} {\enquote {\bibinfo {title} {{Quantum Monte Carlo study of the phase
  diagram of solid molecular hydrogen at extreme pressures}},}\ }\href
  {\doibase doi:10.1038/ncomms8794} {\bibfield  {journal} {\bibinfo  {journal}
  {Nat. Comm.}\ }\textbf {\bibinfo {volume} {6}},\ \bibinfo {pages} {7794}
  (\bibinfo {year} {2015})}\BibitemShut {NoStop}%
\bibitem [{\citenamefont {Wigner}\ and\ \citenamefont
  {Huntington}(1935)}]{Wigner}%
  \BibitemOpen
  \bibfield  {author} {\bibinfo {author} {\bibfnamefont {E.}~\bibnamefont
  {Wigner}}\ and\ \bibinfo {author} {\bibfnamefont {H.~B.}\ \bibnamefont
  {Huntington}},\ }\bibfield  {title} {\enquote {\bibinfo {title} {{On the
  Possibility of a Metallic Modification of Hydrogen}},}\ }\href {\doibase
  http://dx.doi.org/10.1063/1.1749590} {\bibfield  {journal} {\bibinfo
  {journal} {J. Chem. Phys.}\ }\textbf {\bibinfo {volume} {3}},\ \bibinfo
  {pages} {764} (\bibinfo {year} {1935})}\BibitemShut {NoStop}%
\bibitem [{\citenamefont {Eremets}\ and\ \citenamefont
  {Drozdov}()}]{EremetsDrozdov}%
  \BibitemOpen
  \bibfield  {author} {\bibinfo {author} {\bibfnamefont {M.I.}\ \bibnamefont
  {Eremets}}\ and\ \bibinfo {author} {\bibfnamefont {A.~P.}\ \bibnamefont
  {Drozdov}},\ }\bibfield  {title} {\enquote {\bibinfo {title} {{Comments on
  the claimed observation of the Wigner-Huntington Transition to Metallic
  Hydrogen}},}\ }\href {https://arxiv.org/abs/1702.05125} {\bibinfo  {journal}
  {arXiv:1702.05125}\ }\BibitemShut {NoStop}%
\bibitem [{\citenamefont {Mott}(1991)}]{Mott}%
  \BibitemOpen
\bibfield  {journal} {  }\bibfield  {author} {\bibinfo {author} {\bibfnamefont
  {N.~F.}\ \bibnamefont {Mott}},\ }\href@noop {} {\emph {\bibinfo {title}
  {{Metal-Insulator Transitions}}}}\ (\bibinfo  {publisher} {Taylor \& Francis,
  London},\ \bibinfo {year} {1991})\BibitemShut {NoStop}%
\bibitem [{\citenamefont {Gebhard}(1997)}]{Gebhard}%
  \BibitemOpen
  \bibfield  {author} {\bibinfo {author} {\bibfnamefont {F.}~\bibnamefont
  {Gebhard}},\ }\href {http://books.google.de/books?id=uCPgHgEKnwEC} {\emph
  {\bibinfo {title} {{The Mott Metal-Insulator Transition: Models and
  Methods}}}}\ (\bibinfo  {publisher} {Springer},\ \bibinfo {address}
  {Berlin},\ \bibinfo {year} {1997})\BibitemShut {NoStop}%
\bibitem [{\citenamefont {Landau}\ and\ \citenamefont
  {Zeldovich}(1943)}]{Landau}%
  \BibitemOpen
  \bibfield  {author} {\bibinfo {author} {\bibfnamefont {L.~D.}\ \bibnamefont
  {Landau}}\ and\ \bibinfo {author} {\bibfnamefont {Y.~B.}\ \bibnamefont
  {Zeldovich}},\ }\bibfield  {title} {\enquote {\bibinfo {title} {{On the
  relation between the liquid and the gaseous states of metals}},}\ }\href@noop
  {} {\bibfield  {journal} {\bibinfo  {journal} {Acta Physicochimica URSS}\
  }\textbf {\bibinfo {volume} {18}},\ \bibinfo {pages} {194} (\bibinfo {year}
  {1943})}\BibitemShut {NoStop}%
\bibitem [{\citenamefont {Landau}(1969)}]{Landau2}%
  \BibitemOpen
  \bibfield  {author} {\bibinfo {author} {\bibfnamefont {L.~D.}\ \bibnamefont
  {Landau}},\ }\href@noop {} {\enquote {\bibinfo {title} {{Collected works}},}\
  }\bibinfo {howpublished} {Izdatyestvo Nauka, Moscow} (\bibinfo {year}
  {1969}),\ \bibinfo {note} {paper No. 48 (in Russian)}\BibitemShut {NoStop}%
\bibitem [{\citenamefont {Azadi}\ and\ \citenamefont {Foulkes}(2013)}]{Azadi}%
  \BibitemOpen
  \bibfield  {author} {\bibinfo {author} {\bibfnamefont {S.}~\bibnamefont
  {Azadi}}\ and\ \bibinfo {author} {\bibfnamefont {W.~M.~C.}\ \bibnamefont
  {Foulkes}},\ }\bibfield  {title} {\enquote {\bibinfo {title} {{Fate of
  density functional theory in the study of high-pressure solid hydrogen}},}\
  }\href {\doibase 10.1103/PhysRevB.88.014115} {\bibfield  {journal} {\bibinfo
  {journal} {Phys. Rev. B}\ }\textbf {\bibinfo {volume} {88}},\ \bibinfo
  {pages} {014115} (\bibinfo {year} {2013})}\BibitemShut {NoStop}%
\bibitem [{\citenamefont {K\k{a}dzielawa}\ \emph {et~al.}(2015)\citenamefont
  {K\k{a}dzielawa}, \citenamefont {Biborski},\ and\ \citenamefont
  {Spa\l{}ek}}]{Kadzielawa}%
  \BibitemOpen
  \bibfield  {author} {\bibinfo {author} {\bibfnamefont {A.~P.}\ \bibnamefont
  {K\k{a}dzielawa}}, \bibinfo {author} {\bibfnamefont {A.}~\bibnamefont
  {Biborski}}, \ and\ \bibinfo {author} {\bibfnamefont {J.}~\bibnamefont
  {Spa\l{}ek}},\ }\bibfield  {title} {\enquote {\bibinfo {title}
  {{Discontinuous transition of molecular-hydrogen chain to the quasiatomic
  state: Combined exact diagonalization and ab initio approach}},}\ }\href
  {\doibase 10.1103/PhysRevB.92.161101} {\bibfield  {journal} {\bibinfo
  {journal} {Phys. Rev. B}\ }\textbf {\bibinfo {volume} {92}},\ \bibinfo
  {pages} {161101R} (\bibinfo {year} {2015})}\BibitemShut {NoStop}%
\bibitem [{\citenamefont {Biborski}\ \emph {et~al.}(2015)\citenamefont
  {Biborski}, \citenamefont {Kądzielawa},\ and\ \citenamefont
  {Spałek}}]{Biborski}%
  \BibitemOpen
  \bibfield  {author} {\bibinfo {author} {\bibfnamefont {A.}~\bibnamefont
  {Biborski}}, \bibinfo {author} {\bibfnamefont {A.~P.}\ \bibnamefont
  {Kądzielawa}}, \ and\ \bibinfo {author} {\bibfnamefont {J.}~\bibnamefont
  {Spałek}},\ }\bibfield  {title} {\enquote {\bibinfo {title} {{Combined
  shared and distributed memory ab-initio computations of molecular-hydrogen
  systems in the correlated state: process pool solution and two-level
  parallelism}},}\ }\href@noop {} {\bibfield  {journal} {\bibinfo  {journal}
  {Comp. Phys. Commun.}\ }\textbf {\bibinfo {volume} {197}},\ \bibinfo {pages}
  {7} (\bibinfo {year} {2015})}\BibitemShut {NoStop}%
\bibitem [{\citenamefont {K\k{a}dzielawa}\ \emph {et~al.}(2013)\citenamefont
  {K\k{a}dzielawa}, \citenamefont {Spałek}, \citenamefont {Kurzyk},\ and\
  \citenamefont {Wójcik}}]{Kadzielawa2}%
  \BibitemOpen
  \bibfield  {author} {\bibinfo {author} {\bibfnamefont {A.~P.}\ \bibnamefont
  {K\k{a}dzielawa}}, \bibinfo {author} {\bibfnamefont {J.}~\bibnamefont
  {Spałek}}, \bibinfo {author} {\bibfnamefont {J.}~\bibnamefont {Kurzyk}}, \
  and\ \bibinfo {author} {\bibfnamefont {W.}~\bibnamefont {Wójcik}},\
  }\bibfield  {title} {\enquote {\bibinfo {title} {{Extended Hubbard model with
  renormalized Wannier wave functions in the correlated state III}},}\ }\href
  {\doibase 10.1140/epjb/e2013-40127-y} {\bibfield  {journal} {\bibinfo
  {journal} {Eur. Phys. J. B}\ }\textbf {\bibinfo {volume} {86}},\ \bibinfo
  {pages} {252} (\bibinfo {year} {2013})}\BibitemShut {NoStop}%
\bibitem [{\citenamefont {Spa\l{}ek}\ \emph {et~al.}(2000)\citenamefont
  {Spa\l{}ek}, \citenamefont {Podsiad\l{}y}, \citenamefont {W\'ojcik},\ and\
  \citenamefont {Rycerz}}]{Spalek}%
  \BibitemOpen
  \bibfield  {author} {\bibinfo {author} {\bibfnamefont {J.}~\bibnamefont
  {Spa\l{}ek}}, \bibinfo {author} {\bibfnamefont {R.}~\bibnamefont
  {Podsiad\l{}y}}, \bibinfo {author} {\bibfnamefont {W.}~\bibnamefont
  {W\'ojcik}}, \ and\ \bibinfo {author} {\bibfnamefont {A.}~\bibnamefont
  {Rycerz}},\ }\bibfield  {title} {\enquote {\bibinfo {title} {{Optimization of
  single-particle basis for exactly soluble models of correlated electrons}},}\
  }\href {\doibase 10.1103/PhysRevB.61.15676} {\bibfield  {journal} {\bibinfo
  {journal} {Phys. Rev. B}\ }\textbf {\bibinfo {volume} {61}},\ \bibinfo
  {pages} {15676} (\bibinfo {year} {2000})}\BibitemShut {NoStop}%
\bibitem [{\citenamefont {Hubbard}(1963)}]{Hubbard}%
  \BibitemOpen
  \bibfield  {author} {\bibinfo {author} {\bibfnamefont {J.}~\bibnamefont
  {Hubbard}},\ }\bibfield  {title} {\enquote {\bibinfo {title} {{Electron
  Correlations in Narrow Energy Bands}},}\ }\href {\doibase
  10.1098/rspa.1963.0204} {\bibfield  {journal} {\bibinfo  {journal} {Proc.
  Roy. Soc. (London)}\ }\textbf {\bibinfo {volume} {276}},\ \bibinfo {pages}
  {238--257} (\bibinfo {year} {1963})}\BibitemShut {NoStop}%
\bibitem [{\citenamefont {Fazekas}(1999)}]{Fazekas}%
  \BibitemOpen
  \bibfield  {author} {\bibinfo {author} {\bibfnamefont {P.}~\bibnamefont
  {Fazekas}},\ }\href@noop {} {\emph {\bibinfo {title} {{Lecture Notes on
  Electron Correlation and Magnetism}}}}\ (\bibinfo  {publisher} {World
  Scientific, Singapore},\ \bibinfo {year} {1999})\ \bibinfo {note} {chapter 8
  \& 9}\BibitemShut {NoStop}%
\bibitem [{\citenamefont {Fulde}(2012)}]{Fulde}%
  \BibitemOpen
  \bibfield  {author} {\bibinfo {author} {\bibfnamefont {P.}~\bibnamefont
  {Fulde}},\ }\href@noop {} {\emph {\bibinfo {title} {{Correlated Electrons in
  Quantum Matter}}}}\ (\bibinfo  {publisher} {World Scientific, New Jersey},\
  \bibinfo {year} {2012})\BibitemShut {NoStop}%
\bibitem [{\citenamefont {Fetter}\ and\ \citenamefont
  {Walecka}(2003)}]{Fetter}%
  \BibitemOpen
  \bibfield  {author} {\bibinfo {author} {\bibfnamefont {A.~L.}\ \bibnamefont
  {Fetter}}\ and\ \bibinfo {author} {\bibfnamefont {J.~D.}\ \bibnamefont
  {Walecka}},\ }\href {http://www.worldcat.org/isbn/0486428273} {\emph
  {\bibinfo {title} {{Quantum Theory of Many-Particle Systems}}}}\ (\bibinfo
  {publisher} {Dover},\ \bibinfo {year} {2003})\BibitemShut {NoStop}%
\bibitem [{\citenamefont {Hubbard}(1964)}]{Hubbard2}%
  \BibitemOpen
  \bibfield  {author} {\bibinfo {author} {\bibfnamefont {J.}~\bibnamefont
  {Hubbard}},\ }\bibfield  {title} {\enquote {\bibinfo {title} {{Electron
  Correlations in Narrow Energy Bands. III. An Improved Solution}},}\ }\href
  {\doibase 10.1098/rspa.1964.0190} {\bibfield  {journal} {\bibinfo  {journal}
  {Proc. Roy. Soc. (London)}\ }\textbf {\bibinfo {volume} {281}},\ \bibinfo
  {pages} {401} (\bibinfo {year} {1964})}\BibitemShut {NoStop}%
\bibitem [{\citenamefont {Brinkman}\ and\ \citenamefont
  {Rice}(1970)}]{Brinkman}%
  \BibitemOpen
  \bibfield  {author} {\bibinfo {author} {\bibfnamefont {W.~F.}\ \bibnamefont
  {Brinkman}}\ and\ \bibinfo {author} {\bibfnamefont {T.~M.}\ \bibnamefont
  {Rice}},\ }\bibfield  {title} {\enquote {\bibinfo {title} {{Application of
  Gutzwiller's Variational Method to the Metal-Insulator Transition}},}\ }\href
  {\doibase 10.1103/PhysRevB.2.4302} {\bibfield  {journal} {\bibinfo  {journal}
  {Phys. Rev. B}\ }\textbf {\bibinfo {volume} {2}},\ \bibinfo {pages} {4302}
  (\bibinfo {year} {1970})}\BibitemShut {NoStop}%
\bibitem [{\citenamefont {Spałek}\ and\ \citenamefont
  {Wójcik}(1995)}]{Spalek2}%
  \BibitemOpen
  \bibfield  {author} {\bibinfo {author} {\bibfnamefont {J.}~\bibnamefont
  {Spałek}}\ and\ \bibinfo {author} {\bibfnamefont {W.}~\bibnamefont
  {Wójcik}},\ }\bibfield  {title} {\enquote {\bibinfo {title} {{Almost
  Localized Fermions and Mott-Hubbard Transitions at Non-Zero Temperature}},}\
  }in\ \href@noop {} {\emph {\bibinfo {booktitle} {{Spectroscopy of Mott
  Insulators and Correlated Metals}}}},\ \bibinfo {editor} {edited by\ \bibinfo
  {editor} {\bibfnamefont {A.}~\bibnamefont {Fujimore}}\ and\ \bibinfo {editor}
  {\bibfnamefont {Y.}~\bibnamefont {Tokura}}}\ (\bibinfo  {publisher}
  {Springer-Verlag, Berlin},\ \bibinfo {year} {1995})\ \bibinfo {note} {pp.
  41-65}\BibitemShut {NoStop}%
\bibitem [{\citenamefont {Vollhardt}(2014)}]{Vollhardt}%
  \BibitemOpen
  \bibfield  {author} {\bibinfo {author} {\bibfnamefont {D.}~\bibnamefont
  {Vollhardt}},\ }\bibfield  {title} {\enquote {\bibinfo {title} {{From
  Gutzwiller Wave Function to Dynamical-Field Theory}},}\ }in\ \href@noop {}
  {\emph {\bibinfo {booktitle} {{DMFT ar 25: Infinite Dimensions}}}},\ \bibinfo
  {editor} {edited by\ \bibinfo {editor} {\bibfnamefont {E.}~\bibnamefont
  {Pavarini~et al.}}}\ (\bibinfo  {publisher} {Forschungszentrum J\"{u}lich},\
  \bibinfo {year} {2014})\BibitemShut {NoStop}%
\bibitem [{\citenamefont {Spa\l{}ek}\ \emph {et~al.}(1983)\citenamefont
  {Spa\l{}ek}, \citenamefont {Ole\'{s}},\ and\ \citenamefont
  {Honig}}]{Spalek3}%
  \BibitemOpen
  \bibfield  {author} {\bibinfo {author} {\bibfnamefont {J.}~\bibnamefont
  {Spa\l{}ek}}, \bibinfo {author} {\bibfnamefont {A.~M.}\ \bibnamefont
  {Ole\'{s}}}, \ and\ \bibinfo {author} {\bibfnamefont {J.~M.}\ \bibnamefont
  {Honig}},\ }\bibfield  {title} {\enquote {\bibinfo {title} {{Metal-insulator
  transition and local moments in a narrow band: A simple thermodynamic
  theory}},}\ }\href {\doibase 10.1103/PhysRevB.28.6802} {\bibfield  {journal}
  {\bibinfo  {journal} {Phys. Rev. B}\ }\textbf {\bibinfo {volume} {28}},\
  \bibinfo {pages} {6802} (\bibinfo {year} {1983})}\BibitemShut {NoStop}%
\bibitem [{\citenamefont {Mott}(1956)}]{Mott2}%
  \BibitemOpen
  \bibfield  {author} {\bibinfo {author} {\bibfnamefont {N.~F.}\ \bibnamefont
  {Mott}},\ }\bibfield  {title} {\enquote {\bibinfo {title} {{On the Transition
  to Metallic Conduction in Semiconductors}},}\ }\href {\doibase
  10.1139/p56-151} {\bibfield  {journal} {\bibinfo  {journal} {Can. J. Phys.}\
  }\textbf {\bibinfo {volume} {34}},\ \bibinfo {pages} {1356} (\bibinfo {year}
  {1956})}\BibitemShut {NoStop}%
\bibitem [{\citenamefont {Ma\ifmmode~\acute{s}\else
  \'{s}\fi{}ka}(1998)}]{Maska}%
  \BibitemOpen
  \bibfield  {author} {\bibinfo {author} {\bibfnamefont {M.~M.}\ \bibnamefont
  {Ma\ifmmode~\acute{s}\else \'{s}\fi{}ka}},\ }\bibfield  {title} {\enquote
  {\bibinfo {title} {{Ground-state energy of the Hubbard model:
  Cluster-perturbative results}},}\ }\href {\doibase 10.1103/PhysRevB.57.8755}
  {\bibfield  {journal} {\bibinfo  {journal} {Phys. Rev. B}\ }\textbf {\bibinfo
  {volume} {57}},\ \bibinfo {pages} {8755} (\bibinfo {year}
  {1998})}\BibitemShut {NoStop}%
\bibitem [{\citenamefont {Weir}\ \emph {et~al.}(1996)\citenamefont {Weir},
  \citenamefont {Mitchell},\ and\ \citenamefont {Nellis}}]{Weir}%
  \BibitemOpen
  \bibfield  {author} {\bibinfo {author} {\bibfnamefont {S.~T.}\ \bibnamefont
  {Weir}}, \bibinfo {author} {\bibfnamefont {A.~C.}\ \bibnamefont {Mitchell}},
  \ and\ \bibinfo {author} {\bibfnamefont {W.~J.}\ \bibnamefont {Nellis}},\
  }\bibfield  {title} {\enquote {\bibinfo {title} {{Metallization of Fluid
  Molecular Hydrogen at 140 GPa (1.4 Mbar)}},}\ }\href {\doibase
  10.1103/PhysRevLett.76.1860} {\bibfield  {journal} {\bibinfo  {journal}
  {Phys. Rev. Lett.}\ }\textbf {\bibinfo {volume} {76}},\ \bibinfo {pages}
  {1860} (\bibinfo {year} {1996})}\BibitemShut {NoStop}%
\bibitem [{\citenamefont {Tamblyn}\ and\ \citenamefont
  {Bonev}(2010)}]{Tamblyn}%
  \BibitemOpen
  \bibfield  {author} {\bibinfo {author} {\bibfnamefont {I.}~\bibnamefont
  {Tamblyn}}\ and\ \bibinfo {author} {\bibfnamefont {S.~A.}\ \bibnamefont
  {Bonev}},\ }\bibfield  {title} {\enquote {\bibinfo {title} {{Structure and
  Phase Boundaries of Compressed Liquid Hydrogen}},}\ }\href {\doibase
  10.1103/PhysRevLett.104.065702} {\bibfield  {journal} {\bibinfo  {journal}
  {Phys. Rev. Lett.}\ }\textbf {\bibinfo {volume} {104}},\ \bibinfo {pages}
  {065702} (\bibinfo {year} {2010})}\BibitemShut {NoStop}%
\bibitem [{\citenamefont {Moralesa}\ \emph {et~al.}(2010)\citenamefont
  {Moralesa}, \citenamefont {Pierleoni}, \citenamefont {Schweglerd},\ and\
  \citenamefont {Ceperley}}]{Morales}%
  \BibitemOpen
  \bibfield  {author} {\bibinfo {author} {\bibfnamefont {M.~A.}\ \bibnamefont
  {Moralesa}}, \bibinfo {author} {\bibfnamefont {C.}~\bibnamefont {Pierleoni}},
  \bibinfo {author} {\bibfnamefont {E.}~\bibnamefont {Schweglerd}}, \ and\
  \bibinfo {author} {\bibfnamefont {D.~M.}\ \bibnamefont {Ceperley}},\
  }\bibfield  {title} {\enquote {\bibinfo {title} {{Evidence for a first-order
  liquid-liquid transition in high-pressure hydrogen from ab initio
  simulations}},}\ }\href {\doibase 10.1073/pnas.1007309107} {\bibfield
  {journal} {\bibinfo  {journal} {Proc. Natl. Acad. Sci.}\ }\textbf {\bibinfo
  {volume} {107}},\ \bibinfo {pages} {12799} (\bibinfo {year}
  {2010})}\BibitemShut {NoStop}%
\bibitem [{\citenamefont {Silvera}(2010)}]{Silvera}%
  \BibitemOpen
  \bibfield  {author} {\bibinfo {author} {\bibfnamefont {I.~F.}\ \bibnamefont
  {Silvera}},\ }\bibfield  {title} {\enquote {\bibinfo {title} {{The
  insulator-metal transition in hydrogen}},}\ }\href {\doibase
  10.1073/pnas.1007947107} {\bibfield  {journal} {\bibinfo  {journal} {Proc.
  Natl. Acad. Sci.}\ }\textbf {\bibinfo {volume} {107}},\ \bibinfo {pages}
  {12743} (\bibinfo {year} {2010})}\BibitemShut {NoStop}%
\bibitem [{\citenamefont {Nellis}(2015)}]{Nellis2015}%
  \BibitemOpen
  \bibfield  {author} {\bibinfo {author} {\bibfnamefont {W.J.}\ \bibnamefont
  {Nellis}},\ }\bibfield  {title} {\enquote {\bibinfo {title} {{Dynamic high
  pressure: Why it makes metallic fluid hydrogen}},}\ }\href {\doibase
  http://dx.doi.org/10.1016/j.jpcs.2014.12.007} {\bibfield  {journal} {\bibinfo
   {journal} {Journal of Physics and Chemistry of Solids}\ }\textbf {\bibinfo
  {volume} {84}},\ \bibinfo {pages} {49 -- 56} (\bibinfo {year}
  {2015})}\BibitemShut {NoStop}%
\bibitem [{\citenamefont {Eremets}\ and\ \citenamefont
  {Troyan}(2011)}]{Eremets}%
  \BibitemOpen
  \bibfield  {author} {\bibinfo {author} {\bibfnamefont {M.~I.}\ \bibnamefont
  {Eremets}}\ and\ \bibinfo {author} {\bibfnamefont {I.~A.}\ \bibnamefont
  {Troyan}},\ }\bibfield  {title} {\enquote {\bibinfo {title} {{Conductive
  dense hydrogen}},}\ }\href {\doibase 10.1038/nmat3175} {\bibfield  {journal}
  {\bibinfo  {journal} {Nat. Mat}\ }\textbf {\bibinfo {volume} {10}},\ \bibinfo
  {pages} {927} (\bibinfo {year} {2011})}\BibitemShut {NoStop}%
\bibitem [{\citenamefont {Kądzielawa}\ \emph {et~al.}(2014)\citenamefont
  {Kądzielawa}, \citenamefont {Bielas}, \citenamefont {Acquarone},
  \citenamefont {Biborski}, \citenamefont {Maśka},\ and\ \citenamefont
  {Spałek}}]{Kadzielawa3}%
  \BibitemOpen
  \bibfield  {author} {\bibinfo {author} {\bibfnamefont {A.}~\bibnamefont
  {Kądzielawa}}, \bibinfo {author} {\bibfnamefont {A.}~\bibnamefont {Bielas}},
  \bibinfo {author} {\bibfnamefont {M.}~\bibnamefont {Acquarone}}, \bibinfo
  {author} {\bibfnamefont {A.}~\bibnamefont {Biborski}}, \bibinfo {author}
  {\bibfnamefont {M.~M.}\ \bibnamefont {Maśka}}, \ and\ \bibinfo {author}
  {\bibfnamefont {J.}~\bibnamefont {Spałek}},\ }\bibfield  {title} {\enquote
  {\bibinfo {title} {${H}_2$ and $({H}_2)_2$ molecules with an ab initio
  optimization of wave functions in correlated state: electron–proton
  couplings and intermolecular microscopic parameters},}\ }\href
  {http://stacks.iop.org/1367-2630/16/i=12/a=123022} {\bibfield  {journal}
  {\bibinfo  {journal} {New J. Phys.}\ }\textbf {\bibinfo {volume} {16}},\
  \bibinfo {pages} {123022} (\bibinfo {year} {2014})}\BibitemShut {NoStop}%
\bibitem [{\citenamefont {Azadi}\ \emph {et~al.}(2017)\citenamefont {Azadi},
  \citenamefont {Drummond},\ and\ \citenamefont {Foulkes}}]{Azadi2}%
  \BibitemOpen
  \bibfield  {author} {\bibinfo {author} {\bibfnamefont {S.}~\bibnamefont
  {Azadi}}, \bibinfo {author} {\bibfnamefont {N.~D.}\ \bibnamefont {Drummond}},
  \ and\ \bibinfo {author} {\bibfnamefont {W.~M.~C.}\ \bibnamefont {Foulkes}},\
  }\bibfield  {title} {\enquote {\bibinfo {title} {{Nature of the metallization
  transition in solid hydrogen}},}\ }\href {\doibase
  10.1103/PhysRevB.95.035142} {\bibfield  {journal} {\bibinfo  {journal} {Phys.
  Rev. B}\ }\textbf {\bibinfo {volume} {95}},\ \bibinfo {pages} {035142}
  (\bibinfo {year} {2017})}\BibitemShut {NoStop}%
\bibitem [{\citenamefont {McMahon}\ and\ \citenamefont
  {Ceperley}(2011)}]{McMahon}%
  \BibitemOpen
  \bibfield  {author} {\bibinfo {author} {\bibfnamefont {J.~M.}\ \bibnamefont
  {McMahon}}\ and\ \bibinfo {author} {\bibfnamefont {D.~M.}\ \bibnamefont
  {Ceperley}},\ }\bibfield  {title} {\enquote {\bibinfo {title} {{Ground-State
  Structures of Atomic Metallic Hydrogen}},}\ }\href {\doibase
  10.1103/PhysRevLett.106.165302} {\bibfield  {journal} {\bibinfo  {journal}
  {Phys. Rev. Lett.}\ }\textbf {\bibinfo {volume} {106}},\ \bibinfo {pages}
  {165302} (\bibinfo {year} {2011})}\BibitemShut {NoStop}%
\bibitem [{\citenamefont {Kudryashov}\ \emph {et~al.}(2016)\citenamefont
  {Kudryashov}, \citenamefont {Kutukov},\ and\ \citenamefont
  {Mazur}}]{Kudryashov}%
  \BibitemOpen
  \bibfield  {author} {\bibinfo {author} {\bibfnamefont {N.~A.}\ \bibnamefont
  {Kudryashov}}, \bibinfo {author} {\bibfnamefont {A.~A.}\ \bibnamefont
  {Kutukov}}, \ and\ \bibinfo {author} {\bibfnamefont {E.~A.}\ \bibnamefont
  {Mazur}},\ }\bibfield  {title} {\enquote {\bibinfo {title} {{Critical
  Temperature of Metallic Hydrogen at a Pressure of 500 GPa}},}\ }\href@noop {}
  {\bibfield  {journal} {\bibinfo  {journal} {Pis'ma v Zh. Eksp. Teor. Fiz.}\
  }\textbf {\bibinfo {volume} {104}},\ \bibinfo {pages} {488} (\bibinfo {year}
  {2016})},\ \bibinfo {note} {[JETP Lett., \textbf{103}, 460
  (2016)]}\BibitemShut {NoStop}%
\bibitem [{\citenamefont {Spa\l{}ek}\ \emph {et~al.}(2017)\citenamefont
  {Spa\l{}ek}, \citenamefont {Zegrodnik},\ and\ \citenamefont
  {Kaczmarczyk}}]{Spalek4}%
  \BibitemOpen
  \bibfield  {author} {\bibinfo {author} {\bibfnamefont {J.}~\bibnamefont
  {Spa\l{}ek}}, \bibinfo {author} {\bibfnamefont {M.}~\bibnamefont
  {Zegrodnik}}, \ and\ \bibinfo {author} {\bibfnamefont {J.}~\bibnamefont
  {Kaczmarczyk}},\ }\bibfield  {title} {\enquote {\bibinfo {title} {{Universal
  properties of high-temperature superconductors from real-space pairing:
  $t\ensuremath{-}J\ensuremath{-}U$ model and its quantitative comparison with
  experiment}},}\ }\href {\doibase 10.1103/PhysRevB.95.024506} {\bibfield
  {journal} {\bibinfo  {journal} {Phys. Rev. B}\ }\textbf {\bibinfo {volume}
  {95}},\ \bibinfo {pages} {024506} (\bibinfo {year} {2017})}\BibitemShut
  {NoStop}%
\bibitem [{\citenamefont {{Zegrodnik, M. and Spa\l{}ek,
  J.}}(2017)}]{Zegrodnik}%
  \BibitemOpen
  \bibfield  {author} {\bibinfo {author} {\bibnamefont {{Zegrodnik, M. and
  Spa\l{}ek, J.}}},\ }\bibfield  {title} {\enquote {\bibinfo {title} {Effect of
  interlayer processes on the superconducting state within the
  $t\ensuremath{-}j\ensuremath{-}u$ model: Full gutzwiller wave-function
  solution and relation to experiment},}\ }\href {\doibase
  10.1103/PhysRevB.95.024507} {\bibfield  {journal} {\bibinfo  {journal} {Phys.
  Rev. B}\ }\textbf {\bibinfo {volume} {95}},\ \bibinfo {pages} {024507}
  (\bibinfo {year} {2017})}\BibitemShut {NoStop}%
\bibitem [{\citenamefont {Kaczmarczyk}\ \emph {et~al.}(2013)\citenamefont
  {Kaczmarczyk}, \citenamefont {Spa\l{}ek}, \citenamefont {Schickling},\ and\
  \citenamefont {B\"unemann}}]{Kaczmarczyk}%
  \BibitemOpen
  \bibfield  {author} {\bibinfo {author} {\bibfnamefont {J.}~\bibnamefont
  {Kaczmarczyk}}, \bibinfo {author} {\bibfnamefont {J.}~\bibnamefont
  {Spa\l{}ek}}, \bibinfo {author} {\bibfnamefont {T.}~\bibnamefont
  {Schickling}}, \ and\ \bibinfo {author} {\bibfnamefont {J.}~\bibnamefont
  {B\"unemann}},\ }\bibfield  {title} {\enquote {\bibinfo {title}
  {{Superconductivity in the two-dimensional Hubbard model: Gutzwiller wave
  function solution}},}\ }\href {\doibase 10.1103/PhysRevB.88.115127}
  {\bibfield  {journal} {\bibinfo  {journal} {Phys. Rev. B}\ }\textbf {\bibinfo
  {volume} {88}},\ \bibinfo {pages} {115127} (\bibinfo {year}
  {2013})}\BibitemShut {NoStop}%
\bibitem [{\citenamefont {Ashcroft}(1968)}]{Ashcroft}%
  \BibitemOpen
  \bibfield  {author} {\bibinfo {author} {\bibfnamefont {N.~W.}\ \bibnamefont
  {Ashcroft}},\ }\bibfield  {title} {\enquote {\bibinfo {title} {{Metallic
  Hydrogen: A High-Temperature Superconductor?}}}\ }\href {\doibase
  10.1103/PhysRevLett.21.1748} {\bibfield  {journal} {\bibinfo  {journal}
  {Phys. Rev. Lett.}\ }\textbf {\bibinfo {volume} {21}},\ \bibinfo {pages}
  {1748} (\bibinfo {year} {1968})}\BibitemShut {NoStop}%
\bibitem [{\citenamefont {Borinaga}\ \emph {et~al.}(2016)\citenamefont
  {Borinaga}, \citenamefont {Errea}, \citenamefont {Calandra}, \citenamefont
  {Mauri},\ and\ \citenamefont {Bergara}}]{Borinaga}%
  \BibitemOpen
  \bibfield  {author} {\bibinfo {author} {\bibfnamefont {M.}~\bibnamefont
  {Borinaga}}, \bibinfo {author} {\bibfnamefont {I.}~\bibnamefont {Errea}},
  \bibinfo {author} {\bibfnamefont {M.}~\bibnamefont {Calandra}}, \bibinfo
  {author} {\bibfnamefont {F.}~\bibnamefont {Mauri}}, \ and\ \bibinfo {author}
  {\bibfnamefont {A.}~\bibnamefont {Bergara}},\ }\bibfield  {title} {\enquote
  {\bibinfo {title} {{Anharmonic effects in atomic hydrogen: Superconductivity
  and lattice dynamical stability}},}\ }\href {\doibase
  10.1103/PhysRevB.93.174308} {\bibfield  {journal} {\bibinfo  {journal} {Phys.
  Rev. B}\ }\textbf {\bibinfo {volume} {93}},\ \bibinfo {pages} {174308}
  (\bibinfo {year} {2016})}\BibitemShut {NoStop}%
\end{thebibliography}%
\end{document}